\documentclass[10pt,journal,compsoc]{IEEEtran}
\usepackage[colorlinks,
            linkcolor=red,
            anchorcolor=blue,
            citecolor=green
            ]{hyperref}
\usepackage{amsmath}
\usepackage{amsthm}
\usepackage{amssymb}
\usepackage{booktabs}

\usepackage{graphicx}
\usepackage{sidecap}
\usepackage{kantlipsum} %<- For dummy text
\usepackage{mwe} %<- For dummy images
\usepackage{multirow}
\usepackage{multicol}
\newcounter{todocounter}

\usepackage{setspace}
\usepackage{fancyhdr} %%Fancy Kopf- und Fußzeilen
\usepackage{lscape}
\usepackage{rotating} 
\usepackage[autostyle]{csquotes}
\usepackage{amsmath,bm}
\usepackage{array}
\usepackage{tabularx,colortbl} % colored table background
\newcolumntype{C}[1]{>{\centering\arraybackslash}m{#1}}
\newcolumntype{R}[1]{>{\raggedleft\arraybackslash}m{#1}}
\newcolumntype{L}[1]{>{\raggedright\arraybackslash}m{#1}}

\definecolor{slightgray}{gray}{.90} 
\usepackage{rotating}
\usepackage{hhline}
\usepackage{float}
\usepackage{caption}% http://ctan.org/pkg/caption
\newtheoremstyle{mytheoremstyle} % name
        {\topsep}                    % Space above
        {\topsep}                    % Space below
        {\itshape\fontfamily{ptm}\selectfont}                   % Body font, ptm is times new roma
        {}                           % Indent amount
        {\fontfamily{ptm}\selectfont\scshape\color{black}\bfseries}                   % Theorem head font
        {:}                          % Punctuation after theorem head
        {.5em}                       % Space after theorem head
        {}  % Theorem head spec (can be left empty, meaning ‘normal’)

\theoremstyle{mytheoremstyle}
\newtheorem{theorem}{Theorem}[section]

\theoremstyle{mytheoremstyle}

\theoremstyle{mytheoremstyle}

\newtheorem{mydef}{Definition}%[section]

    \makeatletter
    \renewenvironment{proof}[1][\proofname]{%
      \par\pushQED{\qed}\fontfamily{ptm}\selectfont%
      \topsep6\p@\@plus6\p@\relax
      \trivlist\item[\hskip\labelsep\bfseries#1\@addpunct{.}]%
      \ignorespaces
    }{%
      \popQED\endtrivlist\@endpefalse
    }
    \makeatother 
\newcommand{\eg}{e.g., }
\newcommand{\ie}{i.e., }
\usepackage[linesnumbered,ruled,vlined]{algorithm2e}
\usepackage{times}
\usepackage{mathtools}
\usepackage[draft,nomargin,marginclue,footnote,silent]{fixme}
\makeatother
\usepackage{comment}

\usepackage[colorinlistoftodos]{todonotes}

\usepackage{subfigure}
\usepackage{adjustbox}
\usepackage{multirow}
\usepackage{enumitem}
\usepackage{lipsum}
\setlist[itemize]{leftmargin=*}
\usepackage[referable]{threeparttablex}
\renewlist{tablenotes}{enumerate}{1}
\makeatletter
\setlist[tablenotes]{label=\tnote{\alph*},ref=\alph*,itemsep=\z@,topsep=\z@skip,partopsep=\z@skip,parsep=\z@,itemindent=\z@,labelindent=\tabcolsep,labelsep=.2em,leftmargin=*,align=left,before={\footnotesize}}
\makeatother
\usepackage[T1]{fontenc}% optional T1 font encoding

\ifCLASSINFOpdf
\else
\fi
\usepackage{amsmath}
\usepackage{txfonts}
\DeclareMathOperator*{\argmin}{arg\,min}
\DeclareMathOperator*{\argmax}{arg\,max}

\begin{document}
%
% paper title
% Titles are generally capitalized except for words such as a, an, and, as,
% at, but, by, for, in, nor, of, on, or, the, to and up, which are usually
% not capitalized unless they are the first or last word of the title.
% Linebreaks \\ can be used within to get better formatting as desired.
% Do not put math or special symbols in the title.
\title{Versatile Robust Clustering of Ad Hoc Cognitive Radio Network}
%
%
% author names and IEEE memberships
% note positions of commas and nonbreaking spaces ( ~ ) LaTeX will not break
% a structure at a ~ so this keeps an author's name from being broken across
% two lines.
% use \thanks{} to gain access to the first footnote area
% a separate \thanks must be used for each paragraph as LaTeX2e's \thanks
% was not built to handle multiple paragraphs
%

\author{Di~Li,~\IEEEmembership{Member,~IEEE,}
        Erwin~Fang,
        and James~Gross,~\IEEEmembership{Member,~IEEE}% <-this % stops a space
\thanks{D. Li (li@umic.rwth-aachen.de) was with RWTH Aachen university where this work is conducted.}% <-this % stops a space
\thanks{Erwin. Fang was with ETH, Switzerland when this work is conducted.}% <-this % stops a space
\thanks{J. Gross (james.gross@ee.kth.se) is with KTH Royal Institute of Technology, Sweden.}% <-this % stops a space
%\thanks{Manuscript received xxxx xx, 20xx; revised xxxx xx, 20xx.}
}

\maketitle

% As a general rule, do not put math, special symbols or citations
% in the abstract or keywords.
\begin{abstract}
Cluster structure in cognitive radio networks facilitates cooperative spectrum sensing, routing and other functionalities.
The unlicensed channels, which are available for every member of a group of cognitive radio users, consolidate the group into a cluster, and the availability of unlicensed channels decides the robustness of that cluster against the licensed users' influence.
%Thus there should be more unlicensed channels in the clusters.
This paper analyses the problem that how to form robust clusters in cognitive radio network, so that more cognitive radio users can get benefits from cluster structure even when the primary users' operation are intense.
%In the process of forming clusters, every secondary user decides with whom to form a cluster, or which cluster to join.
We provide a formal description of robust clustering problem, prove it to be NP-hard and propose a centralized solution, besides, a distributed solution is proposed to suit the dynamics in the ad hoc cognitive radio network.
Congestion game model is adopted to analyse the process of cluster formation, which not only contributes designing the distributed clustering scheme directly, but also provides the guarantee of convergence into Nash Equilibrium and convergence speed.
Our proposed clustering solution is versatile to fulfill some other requirements such as faster convergence and cluster size control.
The proposed distributed clustering scheme outperforms the related work in terms of cluster robustness, convergence speed and overhead.
%Besides, we prove the clustering problem is NP-hard, and also propose the centralized solution.
The extensive simulation supports our claims.
\end{abstract}

% Note that keywords are not normally used for peerreview papers.
\begin{IEEEkeywords}
cognitive radio, robust cluster, game theory, congestion game, distributed, centralized, cluster size control.
\end{IEEEkeywords}

% For peer review papers, you can put extra information on the cover
% page as needed:
% \ifCLASSOPTIONpeerreview
% \begin{center} \bfseries EDICS Category: 3-BBND \end{center}
% \fi
%
% For peerreview papers, this IEEEtran command inserts a page break and
% creates the second title. It will be ignored for other modes.
\IEEEpeerreviewmaketitle

\graphicspath{
{}
}

\section{Introduction}
\label{intro}
%CR concept
\IEEEPARstart{C}{ognitive} radio (CR) is a promising technology to solve the spectrum scarcity problem~\cite{Mitola}.
Licensed users access the spectrum allocated to them whenever there is information to be transmitted.
In contrast, as one way, unlicensed users can access the spectrum via opportunistic spectrum access, \ie they access the licensed spectrum only after validating the channel is unoccupied by licensed users, where spectrum sensing~\cite{sensing_survey_2009} plays an important role in this process.
In this hierarchical spectrum access model~\cite{zhao_survey_DSA_2007}, the licensed users are also called primary users (PU), while the unlicensed users are referred to as secondary users and constitute a so called cognitive radio network (CRN).
%
%cluster
Regarding the operation of CRN, efficient spectrum sensing is identified to be critical for a smooth operation of a cognitive radio network~\cite{Sahai_FundamentalDesignTradeoffs2006}.
This can be achieved by cooperative spectrum sensing of multiple secondary users, which has been shown to cope effectively with noise uncertainty and channel fading, thus remarkably improving the sensing accuracy~\cite{coorperativeSensing_Akyildiz11}.
Collaborative sensing relies on the consensus of CR users\footnote{The terms user and node appear interchangeably in this paper. In particular, user is adopted when its networking or cognitive ability are discussed or stressed, while we refer node typicallly in the context of the topology.} within a certain area, in this regard, clustering is regarded as an effective method to realize cooperative spectrum sensing~\cite{Sun07_clustering_spectrum_secsing, Zhao07}.
% ... decreases considerably the negative probability due to the random channel effects like fading and shadowing.
Clustering is a process of grouping certain users in a proximity into a collective.
Clustering is also efficient to coordinate the channel switch operation when primary users are detected by at least one CR node residing in the cluster.
The cluster head can enable all the CR devices within the same cluster to stop payload transmission swiftly on the operating channel and to vacate the channel ~\cite{willkomm08}.
%
%As to the coexistence of CR users, they can be notified by cluster head (CH) or other cluster members about the possible collision, the possibility for them to interfere neighbouring clusters is reduced~\cite{centralizedSharing80222}. 
In addition to the collaborate sensing advantage, the use of clusters is beneficial as it reduces the interference between cognitive clusters~\cite{centralizedSharing80222}.
Clustering algorithm has also been proposed to support routing in cognitive radio networks ~\cite{Abbasi_survey_07}.

%%[crn for clustering] 
%%due to attenuation of signal propagation, primary users can only be detected by CR users when they locate closely to CR users.
%In cognitive radio networks, secondary users which locate closely with each other are possibly affected by the same group of primary users, so that the availability of licensed spectrum is similar to them, \ie certain channels are available on each of them.
%The similarity of available spectrum on a group of neighbouring CR nodes, along with the benefit of collaborative decision among multiple nodes, leads to clustering as an effective approach for many applications.
%[robustness issue for clustering]

Clusters are formed in the very beginning of the network operation, and re-formed periodically according to the dynamics of the CRN.
Each formed cluster has one or multiple unlicensed channels which are available for every CR node in the cluster.
The available unlicensed channels are referred to in the following of this paper as \textit{common licensed channels} (or common channels for short, which is abbreviated as CC).
Both payload and control overheads can be transmitted on the CCs.
%Out of the available CCs, there are always one or multiple channels which are used for the payload communication.
When one or several cluster members can not access one certain CC on which primary user activity is detected, the channel will be excluded from the set of CCs.
In particular, if that channel is being used for payload communication, the communication pair will stop and resume the transmission on another available channel from CCs.
The availability of CCs within a cluster defines the existence of that cluster, \ie no CCs are available means the corresponding cluster doesn't exist.
%As long as there is a CC available, the corresponding cluster can continue to exchange payload data.
In the context of CRN, the activity of primary users is usually unknown to the secondary users, thus when the primary users' activity is deemed as random, the cluster which secures more CCs will anticipate a longer time span.
It is obvious that fewer members in a cluster yield more CCs, but obtaining more CCs by decreasing the cluster size contradicts to the motivation of clustering, \ie benefit the cluster members with cooperative decision making.
For example, spectrum sensing accuracy coralates with the cluster size~\cite{Consensus_based_clustering12}, and power consumption doesn't favor small clusters~\cite{clustering_globecom11, EnergyEfficientClusteringRouting_2015}.
In this paper the robustness of clusters means the ability of the clusters to sustain the increasing activity of primary users.
%is synonymous with the affluence of CCs which are possessed by that cluster, to ....

%To solely pursue cluster robustness against the primary users' activity, \ie to achieve more common channels within clusters, the ultimately best clustering strategy is ironically that every node constitutes one cluster (called as \textit{singleton cluster} in this paper).
%%In that case, the common channels within cluster are the available channels available at that node's place.
%Apparently this contradicts our motivation of proposing cluster in cognitive radio network to enable cooperative decision making.
%Thus the cluster robustness discussed in terms of number of CCs carries little meaning when the sizes of formed clusters are not given consideration.

There has been a lot of research done for clustering in wireless networks.
In ad-hoc and mesh networks, the major goal of clustering is to maximize connectivity or to improve the performance of routes~\cite{Kawadia03, clustering_mesh_globecom2010}.
The emphasis of clustering in sensor networks is on network lifetime and coverage~\cite{Abbasi_survey_07}.
%Various clustering schemes are proposed to target different aspects in cognitive radio networks.
%Work~\cite{Consensus_based_clustering12} improves spectrum sensing ability by grouping the CR users with potentially best detection performance into the same cluster.
%Clustering scheme~\cite{clustering_globecom11} obtains the best cluster size which minimizes power consumption caused by communication within and among clusters.
%\cite{clustering_globecom11} proposes clustering strategy in cognitive radio network, which looks into the relationship between cluster size and power consumption and accordingly controlling the cluster size to decrease power consumption.
In respect of CRN, \cite{Zhao07, Chen07,Affinity_clustering_09icccn} propose the clustering schemes to form clusters, where securing CCs is the only goal.
Clustering scheme~\cite{Consensus_based_clustering12} improves spectrum sensing accuracy with cluster structure.
%Clustering scheme~\cite{clustering_globecom11} obtains the best cluster size which minimizes power consumption caused by communication within and among clusters.
\cite{clustering_globecom11, TWC2012_cooperative_communication} target on the QoS provisioning and energy efficiency with cluster structure.
\cite{cluster_EW10} forms clusters to coordinate the control channel usage within one cluster.
An event-driven clustering scheme is proposed for cognitive radio sensor network in \cite{Ozger_cluster_crsn_13}.
No one among the above mentioned schemes provides the robustness to the formed clusters against primary users. 
A clustering scheme (denoted as SOC) which is designed to generate robust clusters against primary users is proposed in~\cite{LIU_TMC11_2}.
SOC involves three phases of distributed executions.
In the first phase, every secondary user forms clusters with some one-hop neighbors, in the second and third phase, each secondary user seeks to either merge other clusters or join one of them.
The metric adopted by every secondary user in the three phases is the product of the number of CCs and cluster size.
The drawbacks are as follows, although the adopted metric considers both cluster size and the number of CCs, cluster formation can be easily dominated by only one factor, e.g. a node which can access many channels may exclude its neighbor and form a cluster by itself.
In addition, this scheme leads to the high variance of the cluster sizes, which is not desired in certain applications as discussed in~\cite{clustering_globecom11, cluster_EW10}.
\cite{mansoor_15_cluster_robust} presents a heuristic method to form clusters, although the authors claim robustness is one goal to achieve, the minimum number of clusters is finally pursued.

A distributed clustering scheme ROSS is proposed in~\cite{Li11_ROSS} under the game theoretic framework. 
Compared with the clustering schemes introduced above, the clusters are formed faster and the clusters possess more CCs within and among clusters than SOC.
But this work doesn't intervene the outcome of singleton clusters, although the cluster sizes are not as divergent as SOC.
Furthermore, this work doesn't consider the robustness of clusters against the increasing activity of primary users, which leaves their claim of robustness unverified.
This paper is on the basis of the work in~\cite{Li11_ROSS}, but extends in two directions.
First, this paper sticks to a new metric of robustness, and considers the clusters when primary users' activity becomes more dynamic.
Second, this paper proposes size control mechanism, which solves the problem of devergence of clusters sizes in ~\cite{Li11_ROSS} and~\cite{LIU_TMC11_2}.
Besides, this paper provides a comprehensive analysis of the robust clustering problem and proposes the centralized solution. 
%We stick to the motivation of forming robust clusters in CRN \ie let more CR users benefited from the cooperative decision making which is due to the clusters. 
%We propose both centralized and distributed clustering schemes, which result in more CR users in the clusters composed with multiple CR users, not only when the clusters are just formed, but also afterwards when the primary users' activity changes.
%Besides, both centralized and distributed schemes take the cluster size into consideration, and both can control the sizes of the formed clusters to certain extent.
%In particular, the decentralized clustering approach ROSS (RObust Spectrum Sharing) is able to form clusters with desired sizes.
%When compared with other distributed works, ROSS involves smaller signaling overhead and the generated clusters are significantly more robust against the primary users which appear after the clusters are formed.
%thesis
%and the generated clusters are more robust than the other clustering scheme which has claims on cluster robustness, \ie more secondary users residing in clusters against increasing influence from primary users.
%
%ROSS selects cluster head through coordination within its neighborhood, and then cluster membership is decided locally and its fast convergence is proved under game theoretic framework. 
%For our scheme we can prove convergence in cluster formation phase and resolve ambiguities with respect to cluster membership in a game-theoretic setting. 
The new extensions are made on basis of ROSS and its light weighted version, the latter involves less overheads thus are more suitable for the scenario where fast deployment is desired.
Throughout this paper, we refer to the clustering schemes on the basis of ROSS as \textit{variants of ROSS}, which include the fast versions and that with size control function.

The rest of paper is organized as follows. 
We present the system model and the robust clustering problem in Section~\ref{sec:model}. 
The centralized and distributed solutions are introduced in Section~\ref{centralized_solution} and~\ref{ross} respectively.
%The clustering problem is given through analysis and a centralized scheme is proposed in section~\ref{centralized_scheme}.
Extensive performance evaluation is presented in Section~\ref{performance}.
Finally, we conclude our work and point out the direction for future research in Section~\ref{conclusion}.

\section{System Model and Problem Formulation}
\label{sec:model}

We consider a set of CR users $\mathcal{N}$ and a set of primary users distributed over a given area.
A set of licensed channels $\mathcal{K}$ is available for the primary users. 
The CR users are allowed to transmit on channel $k \in \mathcal{K}$ only if no primary user is detected on channel $k$. 
CR users conduct spectrum sensing independently and sequentially on all licensed channels.\footnote{We assume that every node can detect the presence of an active primary user on each channel with certain accuracy. The spectrum availability can be validated with a certain probability of detection. Spectrum sensing/validation is out of the scope of this paper.}
We adopt the unit disk model~\cite{unitDiskModel} for both primary and CR users' transmission.
%As to the relation between the primary and secondary users,
If a CR node $i$ locates within the transmission range of an active primary user $p$, then $i$ is not allowed to use the channel which is being used by $p$.
%On the other hand, if $i$ is not in the transmission range of primary user $p$, $i$ can certainly not detect the presence of $p$.
We assume the primary users change their operation channels slowly, then we omit the time index for the spectrum availability, \ie as the result of spectrum sensing, $K_i \subseteq \mathcal{K}$ denotes the set of available licensed channels for CR user $i$ at a time point.
As the transmission range of primary users is limited and secondary users have different locations, different secondary users may have different views of the spectrum availability, i.e., for any $i, j \in \mathcal{N}$, $K_i = K_{j}$ does not necessarily hold.
A cognitive radio network can be represented as a graph $G = (\mathcal{N}, E)$, where $E \subseteq \mathcal{N} \times \mathcal{N}$ such that $\{i, j\} \in E$ if and only if $K_{i} \cap K_{j}\neq \emptyset$ and $d_{i,j} < r$, where $d_{i,j}$ is the distance between $i, j$, $r$ is the radius of secondary user's transmission range. 
Among the secondary users, we denote $\text{Nb}(i)$ as user $i$'s neighborhood, which consists of the CR nodes located within the transmission range of $i$. 

We assume there is one dedicated control channel which is used to exchange signaling messages during clustering process.
This control channel could be one of the ISM bands or other reserved spectrum which is exclusively used for transmitting control messages.\footnote{Actually, the control messages involved in the clustering process can also be transmitted on the available licensed channels through a rendezvous process by channel hopping~\cite{channelHopping_Rendezvous_2014, Gu_distributed_rendezvous_2014}, i.e., two neighboring nodes establish communication on the same channel.}
Over the control channel, a secondary user $i$ can exchange its spectrum sensing result $K_i$ with all its one hop neighbors $\text{Nb}(i)$.
In the following, we refer to the licensed channel as \textit{channel} in general, and will explicitly mention the dedicated control channel if necessary. 

We give the definition of cluster in CRN as follows. 
A cluster $C$ is a set of secondary nodes which possess the same set of CCs.
In particular, one cluster consists of a cluster head $h(C)$ and a number of cluster members, and the cluster head is able to communicate with any cluster member directly.
A cluster can be composed only by the cluster head.
$\text{Nb}(i)$ denotes node $i$'s neighborhood which consists of all its one hop neighbors.
%For any cluster member $i \in C$, $i \in \text{Nb} (h_C) $ holds.
Cluster size of $C$ is written as $|C|$.
Cluster $C(i)$ means the cluster head of this cluster is $i$.
$K(C)$ denotes the set of CCs in cluster $C$, $ K(C) = \cap_{i\in C} K_i$.
%Clustering is performed periodically, because the secondary users are mobile and the primary users change their operation channels, thus the channel availability on secondary users changes accordingly.
The notations used in the system model and the following problem description are listed in Table~\ref{tab1}.
\begin{table}[h!]
\caption{Notations}
\label{tab1}
\centering
\begin{tabular}{llr}
\toprule
%\multicolumn{2}{c}{Item} \\
%\cmidrule(r){1-2}
Symbol & Description \\
\midrule
$\mathcal{N}$  & set of CR users in a CRN\\
$N$ & number of CR users in a CRN, $N=|\mathcal{N}|$\\
$\mathcal{K}$	& set of licensed channels\\
$k(i)$ & the working channel of user $i$\\
$\text{Nb}(i)$ & the neighborhood of CR node $i$    \\
$C(i)$ & a cluster whose cluster head is $i$  \\
%$k_i$   & set of available channels at CR node $i$  \\
$K_i$   & the set of available channels at CR node $i$  \\
$K(C(i))$   & the set of available CCs of cluster $C(i)$ \\
%$h(C)$ & cluster head of a cluster $C$\\
$h(C)$ & the cluster head of a cluster C\\
%$\text{CH}$ & cluster head\\
%$\text{CH}$ & cluster head\\
$\delta$ & the cluster size which is preferred\\
$S_i$ & a set of claiming clusters, each of which includes \\
& debatable node $i$ after phase I\\
$d_i$  & individual connectivity degree of CR node $i$\\
$g_i$  & neighborhood connectivity degree of CR node $i$\\
$f(C)$ & the number of CCs of a cluster $C$, which is used \\
& in the problem description\\
% $\mathcal{G}$ & a collection of some possible clusters in $\mathcal{N}$\\
 $\mathcal{S}$ & the collection of all the possible clusters in $\mathcal{N}$\\
 $C_i$  & the $i$th cluster in $\mathcal{S}$ \\
\bottomrule
\end{tabular}
\end{table}

\subsection{Robust Clustering Problem in CRN}
\label{problem}

As introduced in Section~\ref{intro}, in order to be robust against primary users' activity, the formed clusters should have more CCs to expect longer life expectancy.
On the other hand, the sizes of the formed clusters should not diverge from the desired size greatly.
The formation of small clusters or the \textit{singleton clusters}, \ie the cluster which has only one CR node, contradicts the motivation of forming clusters, as the benefits brought in by the collective of the cluster members are compromised.
On the other hand, large clusters are not preferred in some scenarios neither, \eg for the CRN composed with resource limited users, managing the cluster members in a large cluster is a substantial burden. 
Hence, the cluster size should fall in a desired range according to different application scenarios~\cite{Chen04clusteringalgorithms, capacity_cluster_06}.
Considering the above mentioned requirements, we present the definition of robust clustering problem as follows.

\begin{mydef}
\label{def_centralized_clustering}
\textit{Robust clustering problem in CRN.}

Given a cognitive radio network $\mathcal{N}$ where $|\mathcal{N}|=N$, the collection of all the possible clusters\footnote{Possible cluster means the collection of CR nodes, which complies with the definition in~\ref{sec:model}} in $\mathcal{N}$ is denoted as $\mathcal{S}$ where $\mathcal{S}=\{C_1, C_2,\ldots,C_{|\mathcal{S}|}\}$ \footnote{The subscripts of the clusters can be decided in any convenient way \ie the sequence of identifying them.} and there is $\bigcup_{1\leq i \leq |\mathcal{S}|} C_i = N$.
With the requirements on the cluster size are enforced, \ie the desired size is $\delta$ and the cluster sizes should fall in the scope $\langle\delta_1, \delta_2\rangle$, where $\delta, \delta_1, \delta_2\in \mathbb{Z}^+$, and $\delta_1 \leq \delta\leq \delta_2$, a feasible clustering solution is a subcollection $\mathcal{S}' \subseteq \mathcal{S}$, which satisfies 
$\delta_1\leq|C_i|\leq \delta_2, \bigcup_{C_i\in \mathcal{S}'} C_i = \mathcal{N}$ and $C_{i'}\cap C_i =\emptyset$ where $C_{i'}, C_i\in \mathcal{S}'$ and $i'\neq i$.
The optimal clustering solution is the feasible clustering solution whose sum of the numbers of CCs of clusters $\sum_{C\in \mathcal{S}'} f(C)$ is the maximal.

%To formulate the problem, we propose a new definition for the number of CCs in a cluster.
%We use $f(C_i)$ to denote the number of CCs of a cluster $C_i\in \mathcal{S}$, which is,
%%and the number of common licensed channels is $f(C_i) = |K(C_i)|$ if $|C_i|>1$, and $f(C_i)=0$ when $|C_i|=1$.
%$$
%f(C_i) = \left\{ \begin{array}{rl}
%|K(C_i)| 	&\mbox{$|C_i|>1$} \\
%0 				&\mbox{$|C_i|=1$} \\
%\end{array} \right.
%$$
%According to the new definition of the number of CCs in a cluster, the singleton clusters in $\mathcal{S'}$ don't contribute to the value to be maximized.

\end{mydef}

%The decision version of \textit{weighted exact cover problem}: 
%Given an universe $U$, and collection $S=\{s_1, s_2, \ldots, s_m\}$ where each subset $s_i\subseteq U$ and is given a weight $w_i$, whether there exists a collection of subsets $\mathcal{C}$ and constant number $\lambda$, so that the union of $\mathcal{C}$ equals to $U$, $s_j\cap s_{j'} = \emptyset$ for different $j$ and $j'\in \{1,2,\ldots, m\}$, and $\sum_{i\in J} w_i \geq \lambda$.

\section{Centralized Solution for Robust Clustering}
\label{centralized_solution}

Based on Definition~\ref{def_centralized_clustering}, the decision version of this problem is to determine whether there is a non-empty $\mathcal{S}'\subseteq \mathcal{S}$, so that $\sum_{C\in \mathcal{S}'} f(C) \geqslant \lambda$ where $\lambda$ is a real number.
We have the following theorem on the complexity of this problem.

\begin{theorem}
\label{theorem1}
Robust clustering problem in CRN is NP-hard, when the maximum size of clusters is larger than 3, and $\delta_1=1$ and $\delta_2 = N$.
%Assume a CRN can be represented by a connected graph, and there is at least one common channel between any pair of neighbours, then forming at least two CR nodes into one cluster is NP-complete.
\end{theorem}
The proof is in Appendix~\ref{proof_theorem1}.
%This theorem actually indicates that the robust clustering problem in CRN is NP-hard. 

We propose a centralized solution which solves an optimization with standard solvers. 
To formulate the optimization, we need to do some preparation beforehand.
First, all the possible clusters complying with the description in the system model are found and constitute a set $\mathcal{S}$.
Second, we assign a weight about size to each cluster, which correlates with the difference between the cluster size and the desired cluster size. 
In particular, considering $|\mathcal{S}|=M$, $C_i\in \mathcal{S}$ means the $i$ th cluster in $\mathcal{S}$, and $\delta$ is the desired size, the weight about size for each cluster is given as follows,
$$
p_i(C_i) = \left\{ \begin{array}{rl}
0 &\mbox{ if $|C_i|=\delta$} \\
\rho_1 &\mbox{if $|C_i|=\delta-1$ or $|C_i|=\delta+1$} \\
\rho_2 &\mbox{if $|C_i|=\delta-2$ or $|C_i|=\delta+2$} \\
\vdots
\end{array} \right.
$$
where $\rho_1, \rho_2, \cdots$ are positive values.
In particular, $\rho$ increases with the divergence between $|C_i|$ and $\delta$, \ie these is $\rho_2> \rho_1>0$.

The optimization searches the set $\mathcal{S}$ and decides on certain clusters which constitute the whole CRN without overlapping between any two of them, besides, the sum of CCs of the chosen clusters is maximized.
Then, a central node (or controller) with the knowledge of all CR nodes (also the possible clusters) will solve the problem based on the following formulation. 

%Exact cover problem can be solved with Knuth's Algorithm X~\cite{dancingLinks_Knuth} as it finds out all the instances of exact cover, then we can choose the one with the biggest sum of weights. 
%To make the formulation possible, we adopt heuristics in both the objective function and the constraints, besides, the objective is not totally aligned with the object in the Definition~\ref{def_centralized_clustering}.
%The problem by the Definition~\ref{def_centralized_clustering} is formulate 

%Meanwhile, it provide a chance to constrain the cluster size by putting groups with desired sizes into $\mathcal{C}$.

%the maximum size of $S$ is the \textit{Bell number} of $N$, and $S$ contains the conditions cluster .

%We thus list all possible clusters whose sizes are from 2 to one certain number \footnote{this number of decided by the density of CR network, along with the occupation of PUs. We set this number as cluster size of the biggest cluster ever appears when conducting distributed schemes.}, and check each combination of clusters to find the best covering of network on the aspect of number of ICCs per cluster.
%The complexity of computation is thus \bigO$(N^\delta)$, $\delta$ is the preferred cluster size.

%We apply this centralized scheme on a network with network size $N$ and cluster size $\delta$.
%There is $N\mod \delta=0$, and the expected number of clusters is $C = N/\delta$.
%These tailored parameters don't harm the validity of the performance gap between the two schemes.

\begin{equation}
\begin{aligned}
     &\min\limits_{w_i, x_{ij}} && \Sigma_{j=1}^N\Sigma_{i=1}^M (w_i*t_{ij}) \\
     &\text{subject to}   && \Sigma_{i=1}^G x_{ij} = 1, for \forall j=1, \ldots, N \\
   &&& \Sigma_{j=1}^N x_{ij} = |C_i|*w_i, for \forall i=1, \ldots, M \\
   &&& \text{$x_{ij}$ and $w_i$ are binary variables.}\\
   &&& i\in \{1,2, \cdots M\}, \hspace{0.3cm} j\in \{1,2,\cdots N\}
%\notag
\end{aligned}
\label{centralized_opt}
\end{equation}

%The optimization is a binary linear programming problem.
The objective is to maximize the sum of CCs of all the clusters.
%We minimize the opposite of the real objective to make this problem solvable.
$w_i$ and $x_{ij}$ are the two binary variables in this problem.
% where $j\in \{1,2,\cdots,N-1, N\}$ and $i\in \{1,2,\cdots,M-1, M\}$.
$N$ is the total number of CR users in network $\mathcal{N}$, $M = |\mathcal{S}|$.
%$G = |\mathcal{G}|$ and there is $\mathcal{G} \subseteq \mathcal{S}$.
Being either 1 or 0, $w_i$ denotes whether the $i$th cluster $C_i$ in $\mathcal{S}$ is chosen to be in the solution or not.
$x_{ij}$ indicates whether the CR node $j$ resides in the $i$th potential cluster, \ie $x_{ij}=1$ means node $j$ resides in the cluster $C_i$. 
Node index $j$ is identical to the node ID.
%In practice, we can solve the problem with $\mathcal{G} = \mathcal{S}$ when $|\mathcal{S}|$ is not large. 
%
%
%Given a CRN $\mathcal{N}$ and desired cluster size $\delta$, we obtain a collection of clusters $\mathcal{G}$ which contains all the \textit{potential} clusters, and the sizes of these clusters are $1,2,\ldots,\delta$.
%Potential clusters are the clusters which satisfy the conditions introduced in Section~\ref{sec:model}. 
%Note that the potential clusters include the singleton ones.
%By permitting the existence of the singleton clusters, we can ensure that the $\mathcal{S}'$ in Definition~\ref{def_centralized_clustering} is always be feasible.
$t_{ij}$ is a constant which is $p_i(C_i) - \frac{q_{ij}}{|C_i|}$.
$q_{ij}= |K(C_i)|$ when there is $j\in C_i$, and $q_{ij}= 0$ when there is $j\notin C_i$.
$|K(C_i)|$ is the number of CCs of cluster $C_i$, and $|C_i|$ is the size of cluster $C_i$.
%%Based on the knowledge on $\mathcal{S}$, we construct a $M\times N$ matrix $Q_{M\text{x}N}$ which is shown in Figure~\ref{costant_matrix_Q}. 
%%Constant $q_{ij}$ is the element of $Q_{M\text{x}N}$, and its subscripts correspond to the $i$th cluster and CR node $j$ respectively.
%$q_{ij}= |K(C_i)|/|C_i|$ when there is $j\in C_i$, and $q_{ij}= 0$ when there is $j\notin C_i$.
%In other words, each non-zero element $q_{ij}$ means the number of CCs of the cluster $i$ where node $j$ resides.

%\begin{figure}[ht!]
%\centering
%\bordermatrix{~ 		& 1 	& 2 	& 3 	& \cdots & j & \cdots	& N-1 	& N	\cr
%                  1 	& |K(C_1)| 	& |K(C_1)| 	& 0 	& \cdots & \cdots &\cdots	& 0 	& 0	\cr
%                  2 	& |K(C_2)| 	& 0 	& |K(C_2)| 	& \cdots & \cdots & \cdots 	& 0 	& 0	\cr
%				\vdots  	&\vdots & 	 	& 		&  \vdots		& 		& \vdots \cr
%				i 	& 0 	& |K(C_i)| 	& 0 	& \cdots  & \cdots & \cdots 	& |K(C_i)| 	& 0	\cr
%				\vdots  	&\vdots & 	 	& 		&  \vdots & \cdots & \vdots 		& 		& \vdots \cr
%				\vdots 	& 0  	& 0 	& 0 	& \cdots & \cdots & \cdots 	& |K(C_i')| 	& 0	\cr
%				G  	& |K(C_G)| & \cdots	 	& 		&  \vdots	& \cdots & \vdots& 		& \vdots \cr}	
%\caption{An example of Matrix $Q$, its rows correspond to all possible clusters, and columns correspond to the CR nodes in the CRN. }
%\label{costant_matrix_Q}
%\end{figure}

Now we examine the objective function to see whether it in line with the goal to maximize the total number of CCs meanwhile consider the restriction on cluster size.
The objective function can be written as,
\begin{equation}
\begin{aligned}
     &\min\limits_{w_i, x_{ij}} && \Sigma_{j=1}^N\Sigma_{i=1}^M (-w_i* \frac{q_{ij}}{|C_i|} + w_i*p_i(C_i)) \\
\notag
\end{aligned}
\label{centralized_opt}
\end{equation}
The sum of the first items is the sum of CCs of all the chosen clusters.
The minus sign in front of the first item explains the reason why we minimize instead to maximize the objective function.
As to the second item, when $w_i$ is zero ($C_i$ is chosen), if $|C_i|\neq\delta$, the second component will be positive which contradicts the direction of the optimization.
Thus the second item discourages the appearance the clusters whose sizes are different from $\delta$, especially those whose sizes diverge far from $\delta$.

%$$
%w_i = \left\{ \begin{array}{rl}
%0 &\mbox{if $i$th possible cluster $C_i$ is chosen} \\
%1 &\mbox{if $i$th possible cluster $C_i$ is not chosen} \\
%\end{array} \right.
%$$
%When $w_i$ is zero, it means $C_i$ is chosen by the optimization algorithm, in this case, the second item in the objective function becomes $p$, which is defined as follows,

%Because of $w_i$, any cluster which appears in the clustering solution ($w=0$) will compromise the first component.
%Only when that cluster's size is exactly the preferred size $\delta$, the punishment is zero.
%In contrary, when the chosen cluster's size diverges from $\delta$, the objective function suffers \textit{loss}.
%Then it is easy to notice that when the optimization algorithm adopts singleton clusters, the first xx.
%With such design, the solution is encouraged not to generate singleton clusters.
%doesn't follow the definition of $f(C)$ in Definition~\ref{def_centralized_clustering} strictly, where $f(C_i)=0$ when $|C_i|=1$, but our design echoes the definition by exerting the most severe punishment on the singleton clusters in the clustering solution.
%Choice of $\rho_i$ affects the resultant clusters.

The constraints guarantee to obtain the clusters which together include all the CR users and don't overlap.
The first constraint regulates that each CR node should reside in exactly one cluster.
The second constraint regulates that when the $i$th possible cluster $C_i$ is chosen, there will be exactly $|C_i|$ CR nodes residing in $C_i$.

%\begin{figure}[ht!]
%\centering
%\includegraphics[width=0.45\linewidth]{example.JPG}
%\caption{Example of Matrix Q in a 6-node network, cluster size is set as 2}
%\label{xx}
%\end{figure}

%\begin{figure}[ht!]
%  \centering
%  \includegraphics[width=0.5\linewidth]{figure5final.pdf}
%  \caption{Final cluster formation.}
%  \label{fig4}
%\end{figure}

%Note that the proposed centralized solution is heuristic.
%We reiterate the reasons for pursuing the heuristic scheme, 
%Note the collection of potential clusters is dependant on the network topology and spectrum availability in the network, thus to each specific CRN, the space of solution is different.

This problem is a binary linear programming problem, which can be solved by many available solvers.
The difficulty of using this method lies in the preparation of the set $\mathcal{S}$.
In the worst case \ie the CRN forms a full connected graph, the size of $S$ is $\Sigma_{r=1}^{N}\ {N \choose r} = 2^N-1$.
To levitate this problem, a smaller set \ie $\mathcal{G} \subset \mathcal{S}$ can be used.
$\mathcal{G}$ can be prepared based on the cluster size, and it is recommended to include all the singleton clusters to make sure the availability of feasible solutions.

Another obstacle to apply this centralized scheme is, the centralized entity firstly needs to collect the information from all the CR nodes, then computes the clustering solution and distributes it across the whole network.
This process involves a large number of communication overheads.
In CRN, it is necessary to do clustering again on some occasions.
For example, when a certain amount of CR users move away from their clusters, \ie they loose the direction connection with any member in their previous clusters, or a certain amount of clusters can not be maintained as the CC in these clusters don't exist any longer due to primary users' activity. 
Hence, when the spectrum availability and the CR users' location change frequently, the centralized robust clustering scheme is not suitable for CRN.

%We thus list all possible clusters whose sizes are from 2 to one certain number \footnote{this number of decided by the density of CR network, along with the occupation of PUs. We set this number as cluster size of the biggest cluster ever appears when conducting distributed schemes.}, and check each combination of clusters to find the best covering of network on the aspect of number of ICCs per cluster.
%The complexity of computation is thus \bigO$(N^\delta)$, $\delta$ is the preferred cluster size.

\section{Distributed Clustering Algorithm: ROSS}
\label{ross}

In this section we introduce the distributed clustering scheme ROSS.
With ROSS, CR nodes form clusters based on the proximity of the available spectrum in their neighborhood after a series of interactions with their neighbors.
ROSS consists of two cascaded phases: \textit{cluster formation} and \textit{membership clarification}.
In the first phase, clusters are formed quickly and every CR user becomes either a cluster head or a cluster member.
In the second phase, non-overlapping clusters are formed in a way that the CCs of relevant clusters are mostly increased.

%is based on the spectrum sensing results in its neighbours , then

\subsection{Phase I - Cluster Formation}
\label{phaseI}
We assume that before conducting clustering, spectrum sensing, neighbor discovery and exchange of spectrum availability have been completed, so that every CR node is aware of the available channels on themselves and their neighbors.
In this phase, cluster heads are determined after a series of comparisons with their neighbors. 
Two metrics are proposed to characterize the proximity in terms of available spectrum between CR node $i$ and its neighborhood, which are used in the comparisons to decide on the cluster heads.

%After introducing the connectivity vector, we can proceed to introduce the first phase of algorithm ROSS.

%When a CR user is identified as cluster head, its neighborhood becomes a cluster.

\begin{figure}[ht!]
  \centering
\includegraphics[width=0.7\linewidth]{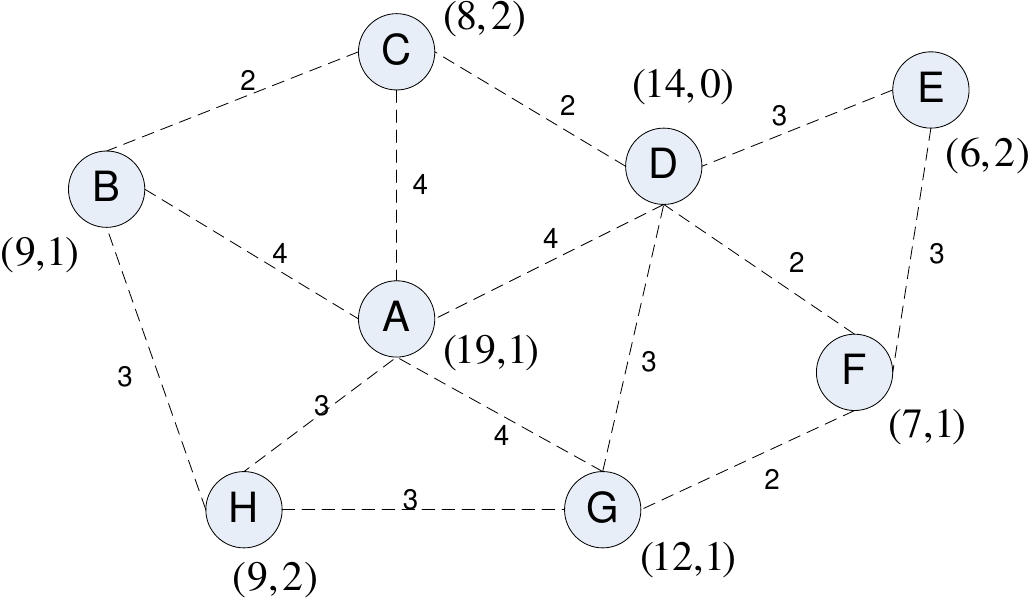}
	\caption{Connectivity graph of the example CRN and the connectivity vector $(d_i, g_i)$ for each node. The desired cluster size $\delta =3$. The sets of the indices of the available channels sensed by each node are: $K_A=\{1,2,3,4,5,6,10\}, K_B=\{1,2,3,5,7\}, K_C=\{1,3,4,10\}, K_D=\{1,2,3,5\}, K_E=\{2,3,5,7\}, K_F=\{2,4,5,6,7\}, K_G=\{1,2,3,4,8\}, K_H=\{1,2,5,8\}$. Dashed edge indicates the end nodes are within each other's transmission range.}
	\label{fig1}
\end{figure}

\begin{itemize}

\item \textit{Individual connectivity degree} $d_i$: $d_i=\sum_{j\in \text{Nb}(i)}\vert K_i\cap K_j\vert$. 
$d_i$ is the total number of the CCs between node $i$ and every its neighbor.
It is an indicator of node $i$'s adhesion to the CRN. 

\item \textit{neighborhood connectivity degree} $g_i$: $g_i=|\bigcap_{j\in \text{Nb}(i)\cup i}K_j|$. 
It is the number of CCs which are available for $i$ and all its neighbors.
$g_i$ represents the ability of $i$ to form a robust cluster with its neighbors.
\end{itemize}
Individual connectivity degree $d_i$ and neighborhood connectivity degree $g_i$ together form the \textit{connectivity vector}.
Figure~\ref{fig1} illustrates an example CRN where every node's connectivity vector is shown.	

\subsubsection{Determining Cluster Heads and Forming Clusters}
The procedure of determining the cluster heads is as follows.
Each CR node decides whether it is a cluster head by comparing its connectivity vector with its neighbors.
When CR node $i$ has lower individual connectivity degree than all its neighbors except for those which have already identified to be cluster heads, node $i$ becomes a cluster head.
%In other words, CR node $i$ becomes cluster head if $d_i > d_k, \forall k\in \texttt{Nb}_i\setminus CHs$ ($CHs$ donate the cluster heads existing in $\texttt{Nb}_i$).
If there is another CR node $j$ in its neighborhood which has the same individual connectivity degree as $i$, \ie $d_j = d_i$ and $d_j < d_{k}, \forall k\in \text{Nb}(j)\setminus \{\Lambda\cup i\}$ where $\Lambda$ denotes the cluster heads, then the node between $i$ and $j$, which has higher neighborhood connectivity degree will become the cluster head, and the other node become one member of the newly identified cluster head. 
If $g_i = g_j$ as well, the node ID is used to break the tie, \ie the one with smaller node ID becomes a cluster head.
The node which is identified as a cluster head broadcasts a message to notify its neighbors of this change, and its neighbors which are not cluster heads become cluster members
\footnote{The reasons for the occurrence of the cluster heads in the neighborhood of a new cluster head will be explained in Section~\ref{ross_p1_guarantee_ccc} and \ref{ross_p2_cluster_pruning})}.
The pseudo code for the cluster head decision and the initial cluster formation is shown in Algorithm~\ref{alg0} in the appendix.

After receiving the notification from a cluster head, a CR node $i$ is aware that it becomes a member of a cluster. 
Consequently, $i$ sets its individual connectivity degree to a positive number $M > |\mathcal{K}| \cdot N$, and broadcasts the new individual connectivity degree to all its neighbors. 
%wrong:
%Note that there is no cluster head which receives the notification on cluster head eligibility, and there is a Lemma on this.
%\begin{theorem}{lemma}
%There is no cluster head receiving the notification message on cluster head eligibility, which is sent from a node which newly becomes a cluster head.
%\end{theorem}
%\begin{proof}
%We use contradiction to prove. 
%Assume a CR node $j$ is cluster head, and receives a notification on cluster head eligibility form CR node $i$. 
%According to the assumption on the reciprocal communication link, $i$ should have received the notification message from $j$ when $j$ becomes cluster head.
%Then CR node $i$ will set its individual connectivity degree to zero, and has no possibility to become a cluster head.
%\end{proof}
When a CR node $i$ is associated to multiple clusters, \ie $i$ has received multiple notifications of cluster head eligibility from different CR nodes, $d_i$ is still set to be $M$. 
%In other words, with this manipulation on its individual connectivity degree, every CR node in the network will be involved in the clustering process, the proof will be given later.
The manipulation of the individual connectivity degree of the cluster members fastens the speed of completing choosing the cluster heads.
%, which enables the nodes which are outside of the cluster can possibly become cluster heads or can also be included into other clusters.
We have the following theorem to show that as long as a secondary user's individual connectivity degree is greater than zero, every secondary user will eventually be either integrated into a certain cluster, or becomes a cluster head.
\begin{theorem}
\label{clustering:theorem}
Given a CRN, it takes at most $N$ steps that every secondary user either becomes cluster head, or gets included into at least one cluster.
\end{theorem}
Here, by \textit{step} we mean one secondary user executing Algorithm~\ref{clustering:theorem} for one time.
The Proof is in Appendix~\ref{proof_clustering:theorem}.

The procedure of the proof also illustrates the time needed to conduct Algorithm~\ref{clustering:theorem}. 
Consider an extreme scenario, where all the secondary nodes sequentially execute Algorithm~\ref{alg0}, \ie they constitute a list as discussed in the example in the proof.
If one step can be finished within certain time $T$, then the total time needed for the network to conduct Algorithm~\ref{clustering:theorem} is $N*T$.
In other scenarios, as Algorithm~\ref{alg0} can be executed concurrently by secondary users which locate in different places, the needed time can be considerably reduced.
%According to Theorem~\ref{clustering:theorem}, Phase I completes within a reasonable amount of time.
%
%This judgement is conducted periodically, and phase I ends after every node ascertains it is cluster head or not.
Let us apply Algorithm~\ref{alg0} to the example shown in Figure~\ref{fig1}.
Node $B$ and $H$ have the same individual connectivity degree, i.e., $d_B=d_H$. As $g_H=2>g_B=1$, node $H$ becomes the cluster head and cluster $C(H)$ is $\{H, B, A, G\}$.

\subsubsection{Guarantee the Existence of Common Channels}
\label{ross_p1_guarantee_ccc}
%After deciding itself being cluster head, CR node broadcasts to notify its neighbours on the dedicated control channel, meanwhile, $i$'s initial cluster is formed immediately, which is $i$'s neighbourhood except for those nodes which have become cluster heads, \ie $C_i=(\text{Nb}_i\setminus \text{CHs})\cup i$.
After executing Algorithm~\ref{alg0}, certain formed clusters may not possess any CCs.
As decreasing cluster size increases the CCs within a cluster, for those clusters having no CCs, certain nodes need to be eliminated to obtain at least one CC.
The sequence of elimination is performed according to an ascending list of nodes which are sorted by the number of common channels between the nodes and the cluster head. 
In other words, the cluster member which has the least common channels with the cluster head is excluded first.
If there are multiple nodes having the same number of common channels with the cluster head, the node whose elimination brings in more common channels will be excluded.
If this criterion meets a tie, the tie will be broken by deleting the node with smaller node ID.
It is possible that the cluster head excludes all its neighbors and resulting in a singleton cluster which is composed by itself.
%At the end of this procedure every formed cluster has at least one common channel.
%The pseudo code for cluster head to obtain at least one common channel is shown in Algorithm~\ref{alg_size_control_available_CCC}.
The pseudo code for this procedure is shown in Algorithm~\ref{alg_size_control_available_CCC}.
As to the nodes which are eliminated from the previous clusters, they restore their original individual connectivity degrees, execute Algorithm~\ref{alg0} and become either cluster heads or get included into other clusters afterwards according to Theorem~\ref{clustering:theorem}.

During Phase I, when ever a CR node is decided to be a cluster head and accordingly forms a cluster, or its cluster's composition is changed, the cluster head will broadcast the updated information about its cluster, which includes the sets of available channels on all its cluster members.

\subsubsection{Cluster Size Control in Dense CRN}
\label{ross_p2_cluster_pruning}

In this subsection, we illustrate the pressing necessity to control the cluster size when CRN becomes denser.

\begin{figure}[h!]
  \centering
  \includegraphics[width=0.4\linewidth]{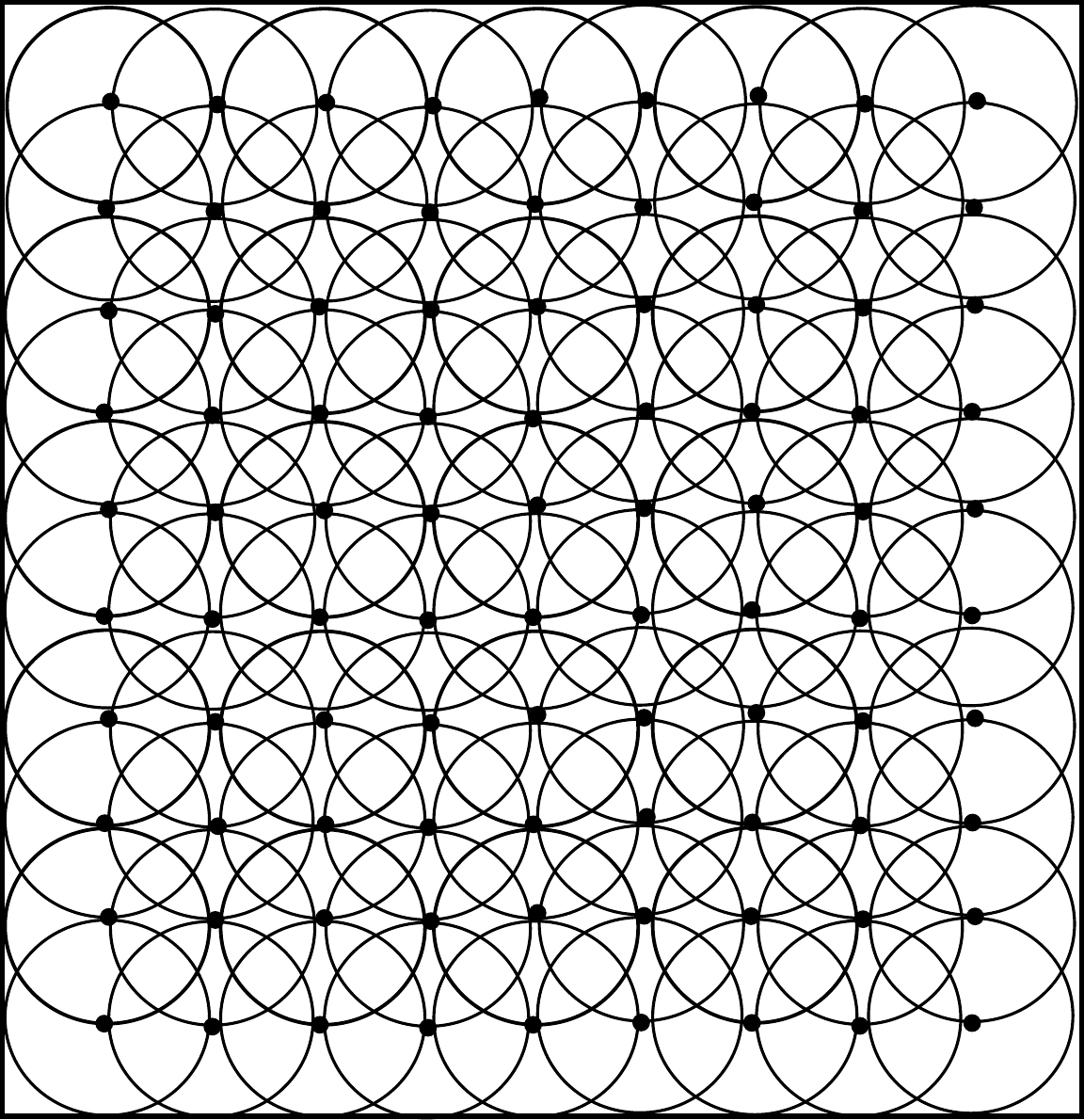}
  \caption{Clusters formation in extremely dense CRN. Black dots are cluster heads, cluster members are not drawn.}
  \label{clusters_denseNetwork}
\end{figure}

\begin{figure}[h]
  \centering
  \includegraphics[width=0.7\linewidth]{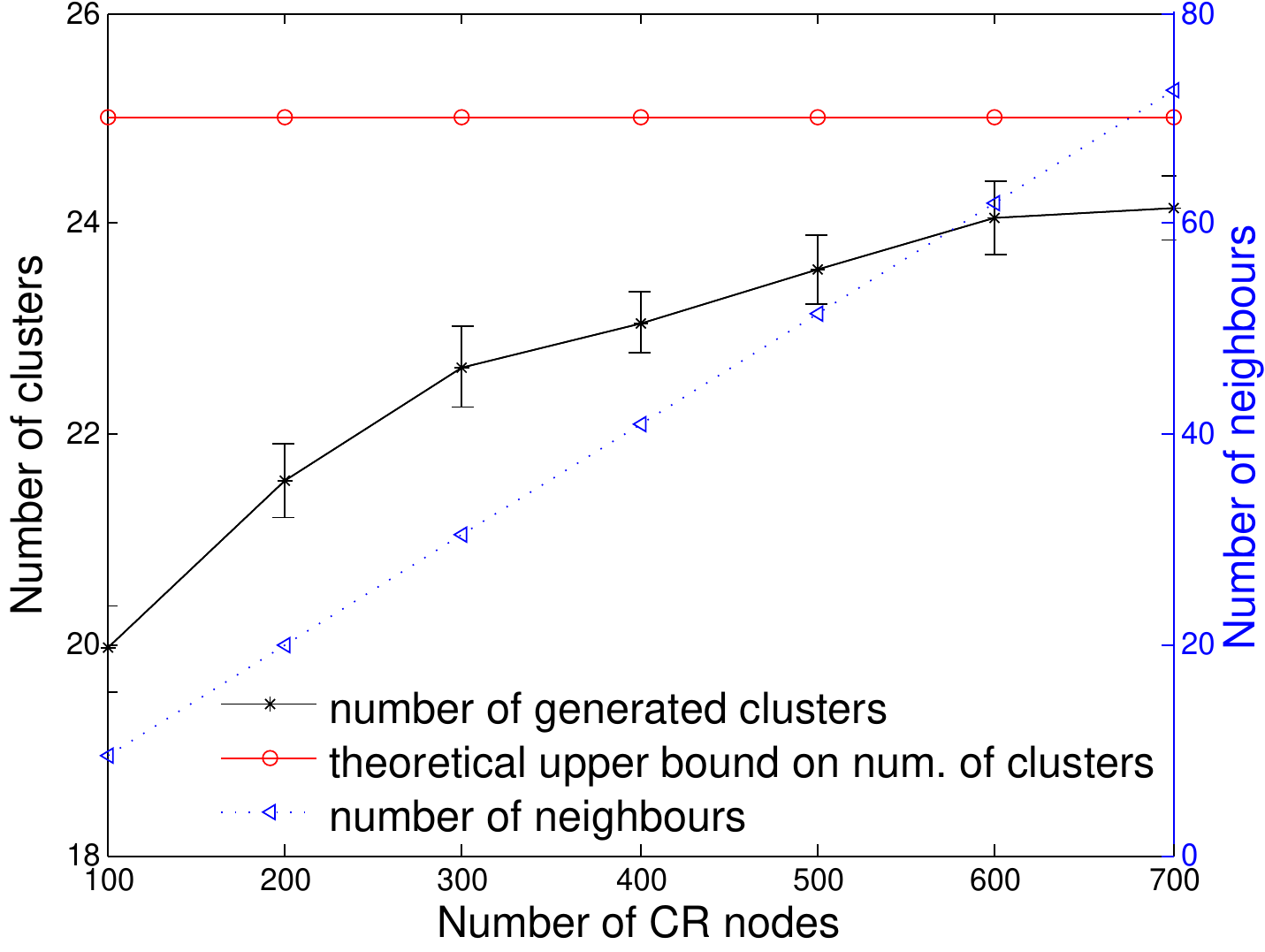}
  \caption{The correlation between the number of formed clusters and network density.}
  \label{number_clusters_scale}
\end{figure}

We consider a cluster $C(i)$ where $i$ is the cluster head in a dense CRN. 
To make the analysis easier, we assume there is no cluster heads which are generated within $i$'s neighborhood during the procedure of guaranteeing CCs.
Assuming the CR users and PUs are evenly distributed and PUs occupy the licensed channels randomly, then both CR nodes density and channel availability in the CRN can be seen to be spatially homogeneous.
In this case the formed clusters are decided by the transmission range and network density.
According to Algorithm~\ref{alg0}, the nearest cluster heads could locate just outside node $i$'s transmission range.
An instance of this situation is shown in Figure~\ref{clusters_denseNetwork}.
In the figure, black dots represent cluster heads, the circles denote the transmission ranges of cluster heads.
Cluster members are not shown in the figure.
%Circles represent the transmission range of cluster head, within which CR nodes are absorbed in cluster.
Let $l$ be the length of side of simulation plan square, and $r$ be CR's transmission radius.
Based on the aforementioned analysis and geometry illustration as shown in Figure~\ref{clusters_denseNetwork}, we give an estimate on the maximum number of generated clusters, which is the product of the number of cluster heads in one row and that number in one line, $l/r * l/r = l^2/r^2$.
%Now we verify the estimation with simulation.
%We distribute CR and primary users randomly on a square plain, $r$=10 and $l$=50.
%Network density is increased by adding more CR users.
%We now have a look at how does the network density affect the cluster size when the transmission range is constant.
%This implicates when the cluster size is decided by the density of the network.
%As to SOC, the membership of one cluster is decided after a complex process, and the cluster size is roughly the same with one neighborhood.xxxx
%We can see from the example that although two neighbouring clusters can overlap greatly with each other, no cluster head will be covered by other clusters.
Given $r$=10 and $l$=50, the maximum number of clusters is 25.
The number of clusters in the simulation is shown in Figure~\ref{number_clusters_scale}.
Simulation is run for 50 times and the confidence interval is 95\%.
With the increase of CR users, network density (the average number of neighbours) increases linearly, and the number of clusters approaches to 25 which complies with the estimation.

Both analysis and simulation show that when applying ROSS, after the clusters are saturated with the increase of network density, the cluster size increases linearly with the network density, thus certain measures are needed to curb this problem.
This task falls to the cluster heads.
To control the cluster size, cluster heads prune their cluster members to reach the desired cluster size.
The desired size $\delta$ is decided based on the capability of the CR users and the tasks to be conveyed.
As there are overlaps between neighboring clusters, the sizes of the clusters formed in this phase are larger than that of the finally formed clusters.
Hence, a cluster head excludes some cluster members when the cluster size exceeds $t\cdot \delta$, where constant parameter $t$ is dependent on the network density and CR nodes' transmission range and $t>1$.
In particular, the cluster head removes the cluster members sequentially according to the following principle, the absence of one cluster member leads to the maximum increase of common channels within the cluster.
This process ends when each cluster's size is smaller or equal to $t \cdot\delta$.
 %delta needs to be changed, f(delta) XXXXXXX
This procedure is similar with guaranteeing the existence of CCs in cluster, thus can reuse Algorithm~\ref{alg_size_control_available_CCC}.
The $t$ is set to 1.3.

\begin{figure}[ht!]
  \centering
  \includegraphics[width=0.5\linewidth]{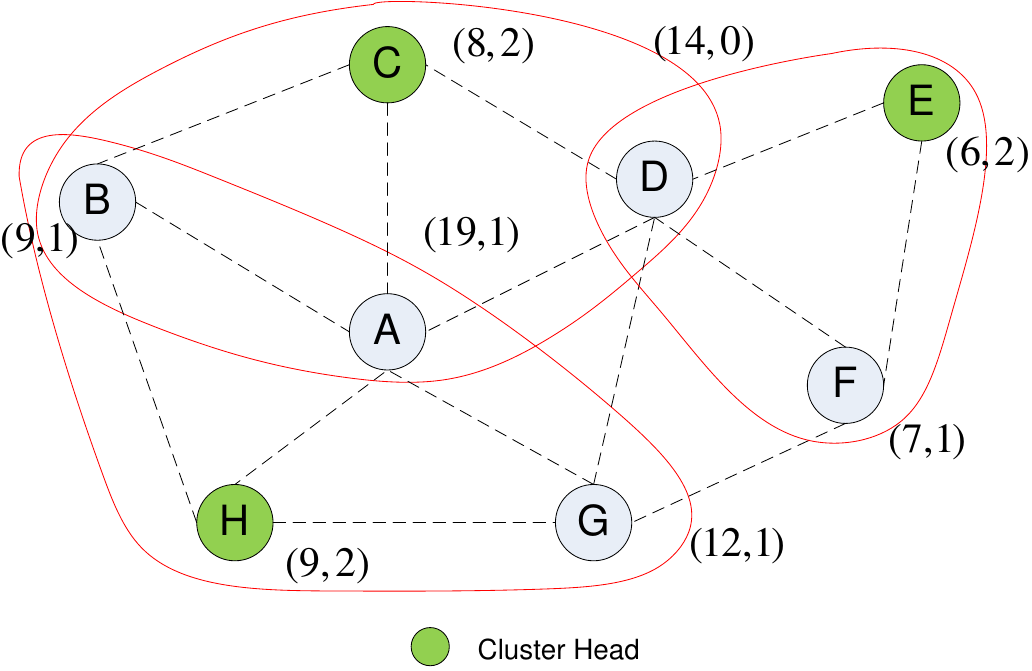}
  \caption{Clusters formation after the phase I of ROSS. CR nodes $A, B, D$ are debatable nodes as they belong to multiple clusters.}
  \label{fig2}
\end{figure}

\subsection{Phase II - Membership Clarification}
\label{membershipClarification}
%\subsubsection*{Problem Description}
As to the example CRN shown in Figure~\ref{fig1}, the resulted clusters are shown in Figure~\ref{fig2} after running phase I of ROSS.
We notice that nodes $A, B, D$ are included in more than one cluster. 
We refer to these nodes as \textit{debatable nodes} as their cluster affiliations are not decided.
The clusters which include the debatable node $i$ are called \textit{claiming clusters} of node $i$, and the set of these clusters is denoted as $S_i$.  
%Actually, debatable nodes extensively exist in CRN with larger scale.
The debatable nodes which are generated from the first phase of ROSS should be exclusively associated with only one cluster and be removed from the other claiming clusters, this procedure is called \textit{cluster membership clarification}.

%In particular, the un-affiliation of one debatable nodes from a cluster increasing the set $K_C$ of ICCs of cluster $C$ (at the cost of potentially decreasing $R_C$, the set of OCCs).

%\newtheorem{observation}{Observation}
%\label{observation}
%\begin{observation}
%If the number of nodes within a cluster decreases, the number of common channels will increase or keep constant.
%\end{observation}
%
%\begin{proof}
%Contradiction, To be continued
%\end{proof}
%From observation 1 we know that the procedure of membership clarification will increase the set of common channels for some clusters and accordingly strengthen the robustness of intra connectivity. 

% % % % %	dependancy!
%An debatable node belongs to multiple clusters, and in the same time, it is possible that several debatable CR nodes locate within one same cluster. %Each debatable node tries to increase the sum of ICC of the clusters which it belong to. More specifically, 
%For a debatable node $i\in S_i$ after phase I, as to clarify its membership, it will choose one cluster $C\in S_i$ to stay and withdraw from the other clusters in $S_i$ with the consideration of increasing ICCs within $S_i$ by the largest margin. 

\subsubsection{Distributed Greedy Algorithm (DGA)}
%Debatable node $i$ is aware of all its claiming clusters in $S_i$. 
Assume a debatable node $i$ needs to decide one cluster $C\in S_i$ to stay, and thereafter leaves the rest others in $S_i$.
In this process, the principle for $i$ is that its move should result in the greatest increase of CCs in all its claiming clusters.
%The set of available channels of one cluster are known by the debatable nodes which locate in that cluster. %this is finished 
Note that node $i$ is aware of the spectrum availability on all the cluster members of each claiming cluster, thus node $i$ is able to calculate how many more CCs can be produced in one claiming cluster if $i$ leaves that cluster.
%Based on this calculation, $i$ decides on one claiming cluster to stay and leaves the other claiming clusters.
If there exists one cluster $C\in S_i$, when $i$ leaves this cluster brings the least increased CCs than leaving any other claiming clusters, then $i$ chooses to stay in cluster $C$.
When there comes a tie, among the claiming clusters, $i$ chooses to stay in the cluster whose cluster head shares the most CCs with $i$.
In case there are multiple claiming clusters demonstrating the same on the aforementioned metric, node $i$ chooses to stay in the claiming cluster which has the smallest size.
Node IDs of cluster heads will be used to break tie if all the previous metrics could not decide on the unique claiming cluster for $i$ to stay.
The pseudo code of this algorithm is given as Algorithm~\ref{alg4}.
%To conduct Algorithm~\ref{alg4}, debatable node $i$ needs to know the necessary information about its claiming clusters, \ie $K(C)$ (the set of available channels in cluster $C$), $K(h_C)$ (the set of available channels on $C$'s cluster head $h_C$) and $|C|, C\in S_i$ (sizes of $i$'s claiming clusters).
After deciding its membership, debatable node $i$ notifies all its claiming clusters of its choice, and the claiming clusters from which node $i$ leaves also broadcast their new cluster composition and the spectrum availability on all their cluster members.

The autonomous decisions made by the debatable CR nodes raise the concern on the endless chain effect in the membership clarification phase.
A debatable node's choice is dependent on the compositions of its claiming clusters, which can be changed by other debatable nodes' decisions.
As a result, the debatable node which makes decision first may change its original choice, and this process may go on forever.
To erase this concern, we formulate the process of membership clarification into a game, where a equilibrium is reached after a finite number of best response updates made by the debatable nodes.

\subsubsection{Bridging ROSS-DGA with Congestion Game}
\label{clustering:phaseII:game}
Game theory is a powerful mathematical tool for studying, modelling and analysing the interactions among individuals.
A game consists of three elements: a set of players, a selfish utility for each player, and a set of feasible strategy space for each player. In a game, the players are rational and intelligent decision makers, which are related with one explicit formalized incentive expression (the utility or cost).
Game theory provides standard procedures to study its equilibriums~\cite{game_for_communication_01}.
In the past few years, game theory has been extensively applied to problems in communication and networking~\cite{Neel06analysisand, Wang_gtc_crn_survey_2010}.
Congestion game is an attractive game model which describes the problem where participants compete for limited resources in a non-cooperative manner, it has good property that Nash equilibrium can be achieved after finite steps of best response dynamic, \ie each player choose strategy to maximizes/minimizes its utility/cost with respect to the other players' strategies.
Congestion game has been used to model certain problems in internet-centric applications or cloud computing, where self-interested clients compete for the centralized resources and meanwhile interact with each other.
For example, server selection is involved in distributed computing platforms~\cite{Cloud_Computing_2010}, or users downloading files from cloud, etc.

\begin{figure}[ht!]
  \centering
  \includegraphics[width=0.25\linewidth]{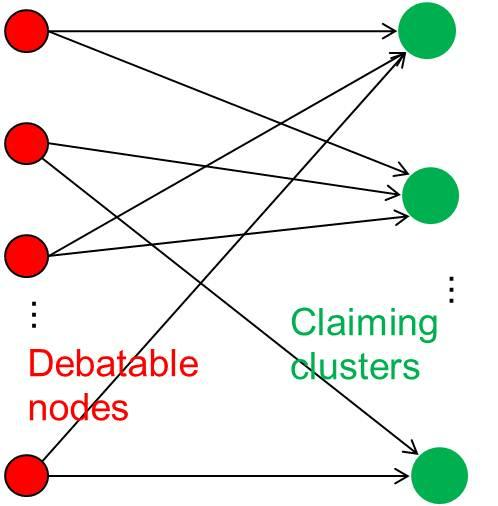}
  \caption{Debatable nodes and claiming clusters}
  \label{debatable_nodes_claiming_cluster}
\end{figure}

To formulate the debatable nodes' membership clarification into the desired congestion game, we observe this process from a different (or opposite) perspective. 
From the new perspective, the debatable nodes are regarded to be isolated and don't belong to any cluster, in other words, their claiming clusters become clusters which are beside them. 
Now for the debatable nodes, the previous problem of deciding which clusters to leave becomes a new problem that which cluster to join.
In the new problem, debatable node $i$ chooses one cluster $C$ out of $S_i$ to join if the decrease of CCs in cluster $C$ is the smallest in $S_i$, and the decrease of CCs in cluster $C$ is $\sum_{C\in S_i}\Delta\vert K(C) \vert=\sum_{C\in S_i}({\vert K(C) \vert-\vert K(C\cup i) \vert})$.
The interaction between the debatable nodes and the claiming clusters is shown in Figure~\ref{debatable_nodes_claiming_cluster}.
%The concern on convergence appears again as we have discussed in the previous subsubsection.
%We give proof on convergence under game theoretic framework.

%In the following we will introduce an \textit{server matching}~\cite{kothari:congestion_serverMatching} problem to illustrate congestion game's application in communication systems.

In the following, we show that the decision of debatable nodes to clarify their membership can be mapped to the behaviour of the players in a \textit{player-specific singleton congestion game} when proper cost function is given.
The game to be constructed is represented with a 4-tuple $\Gamma=(\mathcal{P},\mathcal{R},\sum_{i, i \in \mathcal{P}}, f)$, and the elements in $\Gamma$ are explained below,
%To make the model of this game more clear, we make some change to our original problem. Previously, the nodes in overlapping areas belong to more than one cluster, and our scheme is to remove them out of some clusters to increase the set of common channels within the cluster form which the mode leave. In the new model, we assume all the nodes in overlapping nodes don't belong any cluster and the problem become into how do these nodes decide which cluster to join.

%The components of the game are listed as follows,

\begin{itemize}
%	\item $\mathcal{D}=\left\{1,\ldots,n\right\}$, the set of players (debatable nodes).
	\item $\mathcal{P}$, the set of players in the game, which are the debatable nodes in our problem.
%	\item $\mathcal{R}=\left\{1,\ldots,m\right\}$, the set of resources which player can choose, which are all the clusters in our model.
	\item $\mathcal{R} = \cup S_i, i\in \mathcal{P}$, denotes the set of resources for players to choose, in our problem, $S_i$ is the set of claiming clusters of node $i$, and $\mathcal{R}$ is the set of all claiming clusters.
	\item Strategy space $\sum_i, i \in \mathcal{P}$ is the set of claiming clusters $S_i$.
	As debatable node $i$ is supposed to choose only one claiming cluster, then only one piece of resource will be allocated to $i$.%, accordingly this congestion game is a singleton game.
	%when $i$ makes decision, only one resource (one claiming cluster) from the allowed resources is allocated.
	
	%\item We denote by $\mathcal{S}=\left(\mathcal{S}_1,\ldots,\mathcal{S}_n\right)\in \sum_1\times \cdots\times\sum_n$ the state of game where player $i$ plays strategy $\mathcal{S}_i\in \Sigma_i$.
	
	\item %For the clusters which are possible destination of debatable nodes, the decrease of common channels caused by different debatable node' join can be different because of the heterogeneity of channel availability within itself and on the debatable nodes. %furthermore, the sequence of debatable node's join can also alter the decrease. 
	The utility (cost) function $f(C)$ as to a resource $C$. 
	$f(C) = \Delta\vert K^i(C)|, C\in S_i$, which represents the decreased number of CCs in cluster $C$ when debatable node $i$ joins $C$.
	As to cluster $C\in S_i$, the decrease of CCs caused by including the debatable nodes is $\sum_{i:C\in S_i, i\rightarrow C} \Delta\vert K^i(C) \vert$. 
$i\rightarrow C$ means $i$ joins cluster $C$.
Obviously this function is non-decreasing with respect to the number of nodes joining cluster $C$.
	
The utility function $f$ is not purely decided by the number of players accessing the resource (debatable nodes join claiming clusters), which happens in a canonical congestion game.
The reason is in this game the channel availability on debatable nodes is different.
Given two same groups of debatable nodes and their sizes are the same, when the nodes are not completely the same (neither are the channel availabilities on these nodes), the cost happened on one claiming cluster could be different if the two groups of debatable nodes join that cluster respectively.
%In a canonical congestion game, the cost (or pay off) is function of only the number of players occupying the resource, and is mono-
%In this new game, the cost function 
Hence, this congestion game is player specific~\cite{Ackermann06purenash}.
In this game, every player greedily updates its strategy (choosing one claiming cluster to join) if joining a different claiming cluster minimizes the decrease of CCs $\sum_{i:C\in S_i} \Delta\vert K^i(C) \vert$, and a player's strategy in the game is exactly the same with the behaviour of a debatable node in the membership clarification phas.

%	\item The Rosenthal's potential function \cite{Rosenthal} of this congestion game is given by:
%	\begin{equation*}
%	\phi(S)=\sum_{C\in\mathcal{R}} \sum_{i:C\in S_i} \Delta\vert K^i_C \vert   	
%   	%\sum_{i=1}^N \Delta^{i}_{p}(S)=\sum_{i=1}^N \sum_{r\in S_i}\Delta^{i}_{r}(t)	
%  	% \Delta =\sum_{i=1}^N w_i (x_i - \bar{x})^2 .
%	\end{equation*}
%All the players in this game greedily update their strategy to minimize the potential function (congestion), this process is exactly the same with the network behaviour under \textit{Distributed Greedy Algorithm}. 

%	\item It is an asymmetric game because the sets of strategies shared by different players are different.
%	\item The total cost is: 
%\begin{equation*}
%   \sum_{i=1}^N \Delta^{i}_{p}(S)=\sum_{i=1}^N \sum_{p\in s_i} \Delta^{i}_{p}(n_p(S))
%  % \Delta =\sum_{i=1}^N w_i (x_i - \bar{x})^2 .
%\end{equation*}

%This is the global objective we want to minimize.
\end{itemize}

%Singleton congestion game is a special type of matroid game~\cite{Milchtaich1996111,}. 
%It is known that player-specific matroid congestion game admit pure equilibrium, 

As to singleton congestion game, there exists a pure equilibria which can be reached with the best response update, and the upper bound for the number of steps before convergence is $n^2*m$~\cite{Ackermann06purenash}, where $n$ is the number of players, and $m$ is the number of resources.
In our problem, the players are the debatable nodes, and the resources are the claiming clusters.
Thus the upper bound of the number of steps can be expressed as $\mathcal{O}(N^3)$.

In fact, the number of steps which are actually involved in this process is much smaller than $N^3$, as both $n$ and $m$ are considerably smaller than $N$.
The percentage of debatable nodes in $\mathcal{N}$ is illustrated in Figure~\ref{percentage_overlapping_node}, which is between 10\% to 60\% of the total number of CR nodes in the network.
The number of clusters heads, as discussed in Section~\ref{phaseI}, is dependent on the network density and the CR node's transmission range.
As shown in Figure~\ref{number_clusters_scale}, the cluster heads take up only 3.4\% to 20\% of the total number of CR nodes.

%and the number of steps towards \textit{Nash Equilibrium} is upper-bounded\footnote{Here we present this with modifying the original conclusion in \cite{Ackermann06purenash} according to our model.} by $ n^2\cdot m $. In our context, $n$ is the number of debatable nodes, $m$ is number of clusters in CRN, %$rk(\Gamma)$ of the matroid  is the cardinality of the maximal independent sets, which is 1 in the case of singleton game, 
%so the total time complexity to achieve the \textit{Nash Equilibrium} using greedy approach is 	.
%%(XXX Just mention after this complexity result the relationship to the system mdeol XX)
%This is upper-bounded (in the worst case) by $O(\vert I\vert^3)$. 
%Based on above model and analysis, phase II converges is Algorithm~\ref{alg4} is run by debatable nodes. 
%Although the game version of DGA can achieve \textit{Nash Equilibrium}, the whole scheme can possibly obtain sub-optimal result.    %, furthermore,this stable state is a local minimum of the global decrease function.
%\todo[inline]{The number of steps, or the upper bound of steps in convergence needs a formal proof}

\subsubsection{Distributed Fast Algorithm (DFA)}
%The convergence speed of DGA is large recalling that the number of steps is of $\mathcal{O}(N^3)$.
On the basis of ROSS-DGA, we propose a faster version ROSS-DFA which differs from ROSS-DGA in the second phase.
With ROSS-DFA, debatable nodes decide their respective cluster heads once.
The debatable nodes consider their claiming clusters to include all their debatable nodes, thus the membership of claiming clusters is static and all the debatable nodes can make decision simultaneously without considering the change of membership of their claiming clusters.
As ROSS-DFA is quicker than ROSS-DGA, the former is especially suitable for the CRN where the channel availability changes dynamically and re-clustering is necessary.
To run ROSS-DFA, debatable node executes only one loop in Algorithm~\ref{alg4}.

\begin{figure}[h]
\centering
\subfigure[Node A stays in cluster $C(C)$, quits $C(H)$, $\Delta\vert K(C(C))\vert+\Delta\vert K(C(H))\vert=2$]{
\includegraphics[width=0.435\linewidth]{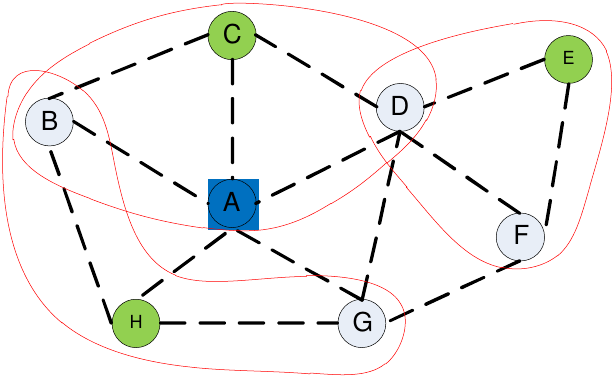}
\label{AinC}
}
\hspace{.15 in}
\subfigure[Node A stays in cluster $C(H)$, quits $C(C)$, $\Delta\vert K(C(C))\vert+\Delta\vert K(C(H))\vert=1$]{
\includegraphics[width=0.435\linewidth]{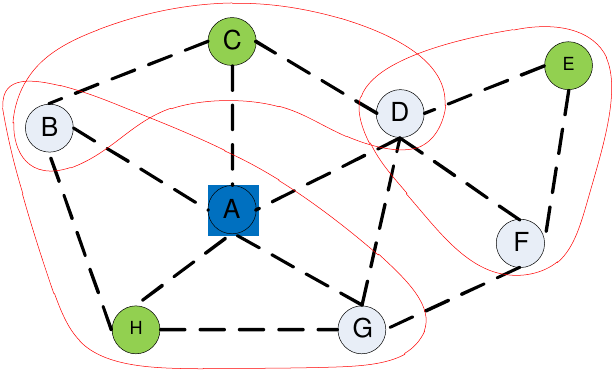}
\label{AinH}
}
\caption[]{Membership clarification: possible cluster formations caused by node A's different choices} %\subref{node A in $C_C$}, \subref{node A in $C_H$}}
\label{fig3}
\end{figure}

\begin{figure}[h]
  \centering
  \includegraphics[width=0.5\linewidth]{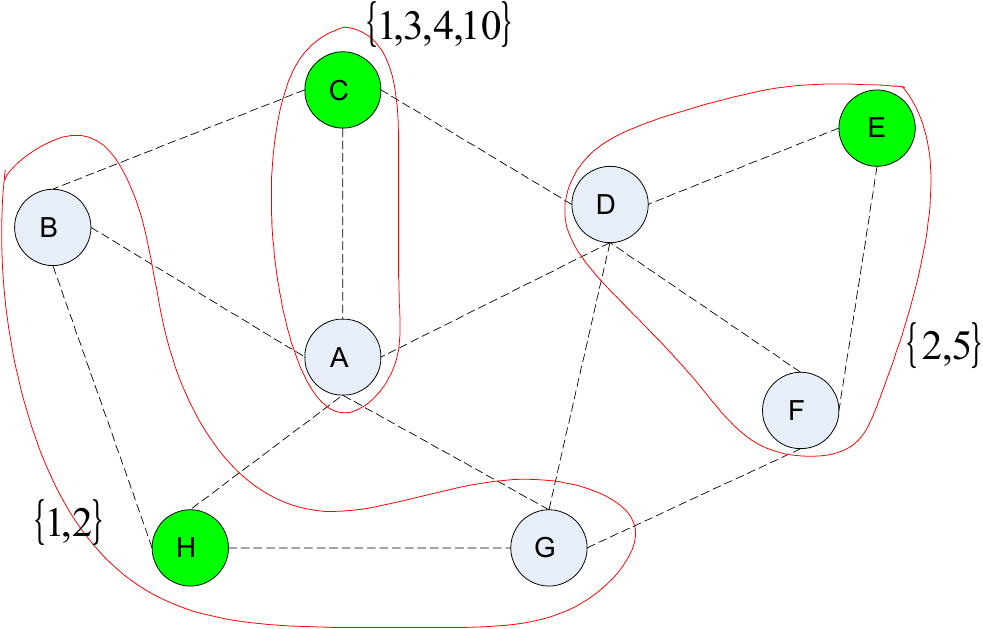}
  \caption{Final formation of clusters. Common channels are shown beside corresponding clusters.}
  \label{final_clustering_ross}
\end{figure}

Now we apply both ROSS-DGA and ROSS-DFA to the toy network in Figure~\ref{fig2} which has been applied the phase I of ROSS.
In the network, node $A$'s claiming clusters are cluster $C(C), C(H)\in S_A$, their members are $\{A,B,C,D\}$ and $\{A,B,H,G\}$ respectively. 
The two possible strategies of node $A$ is illustrated in Figure \ref{fig3}.
In Figure \ref{AinC}, node $A$ staying in $C(C)$ and leaving $C(H)$ brings 2 more CCs to $S_A$, which is more than that brought by another strategy showed in \ref{AinH}.
After the decisions made similarly by the other debatable nodes $B$ and $D$, the final clusters are formed as shown in Figure~\ref{final_clustering_ross}.

%Using DFA in phase II, the time complexity is decreased drastically to 1. Thus, the total complexity of ROSS-DFA is $|I|$, while, ROSS-DGA's complexity is $|I|^3$ in the worst case.

\section{Performance Evaluation}
\label{performance}
%In this section, we evaluate the performances of ROSS.
The schemes involved in the simulation are listed as follows,
\begin{itemize}
\item ROSS without size control, \ie ROSS-DGA and ROSS-DFA.
\item ROSS with size control. \ie ROSS-$\delta$-DGA and ROSS-$\delta$-DFA where $\delta$ is the desired cluster size.
In the following, we refer to the above mentioned four schemes as the variants of ROSS.
\item SOC~\cite{LIU_TMC11_2}, a distributed clustering scheme pursuing cluster robustness.
\item Centralized robust clustering scheme. 
The formulated optimization is an integer linear optimization problem, which is solved by MATLAB with the function $bintprog$.
\end{itemize}

The ROSS without size control mechanism is similar with the schemes proposed in \cite{Li11_ROSS}.
The authors of~\cite{LIU_TMC11_2} compared SOC with other schemes in terms of the average number of CCs of the formed cluster, on which SOC outperforms other schemes by 50\%-100\%. 
SOC's comparison schemes are designed either for ad hoc network without consideration of channel availability~\cite{Basagni99}, or for CRN but just considering connection among CR nodes~\cite{Zhao07}. 
Thus SOC is chosen to be the only distributed scheme as comparison, besides, we also compare ROSS with the centralized scheme.

\begin{figure}[ht]
\begin{center}
\subfigure[Generated by SOC]{\includegraphics[width=0.435\linewidth]{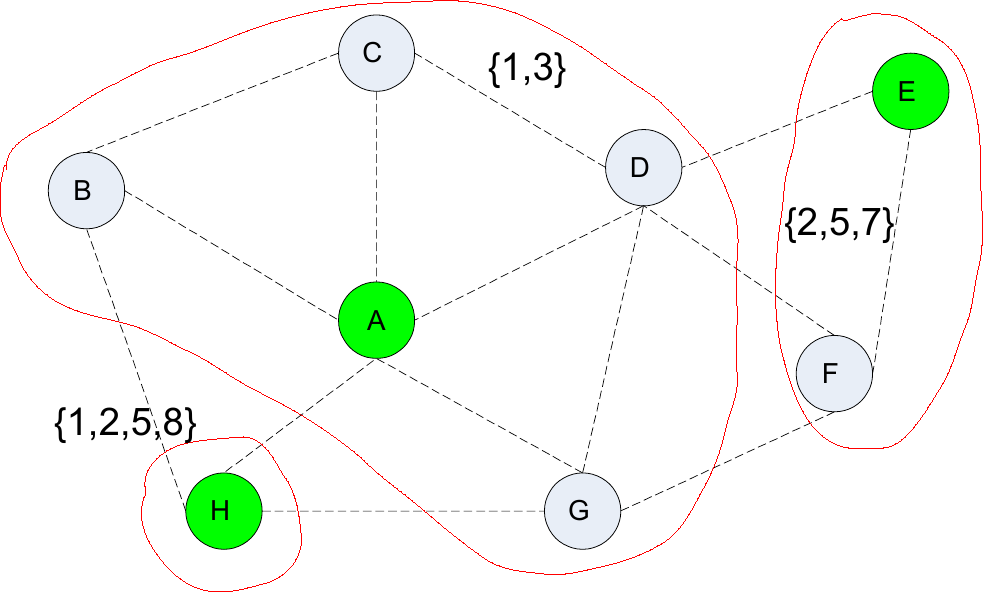}\label{fig:final_clustering_soc}}
\hspace{0.15 in}
\subfigure[Generated by the centralized clustering scheme]{\includegraphics[width=0.435\linewidth]{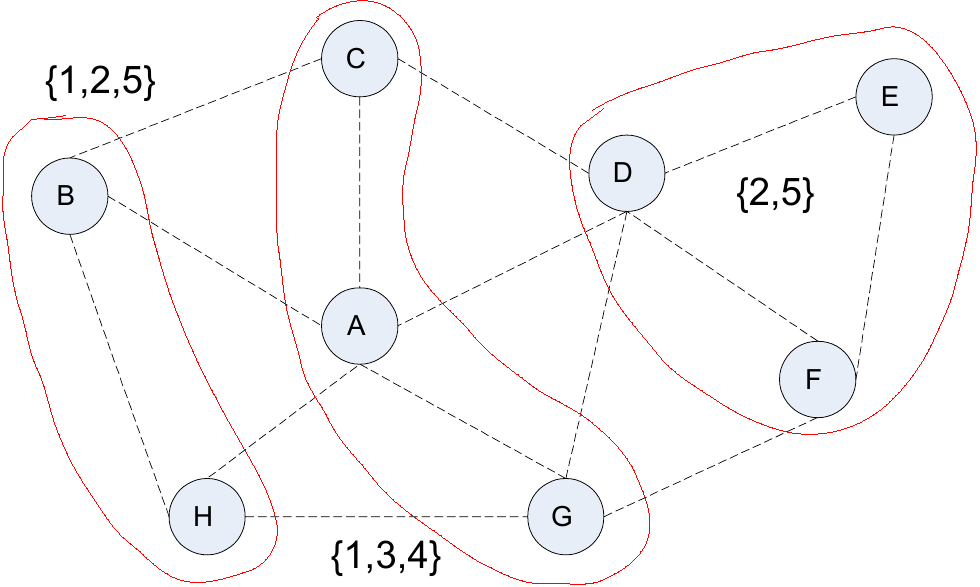}\label{fig:final_clustering_LP}}
\end{center}
\caption{Final clusters formed by the centralized clustering scheme and SOC.}
\label{fig:final_clustering}
\end{figure}

Before we investigating the performance of the clustering schemes with simulation, we apply the two comparison clustering schemes in the example CRN in Figure~\ref{fig1}, and make an initial comparison in terms of the amount of CCs. 
As to the centralized robust clustering scheme, we set the desired cluster size $\delta$ as 3, as a result, according to the network topology, the collection of all the possible clusters
$\mathcal{S}=\{\{A\}, \{B\},\dots,\{B,C\},\{B,A\},\{B,H\},\cdots,\{B,A,C\},$\\$\{B,H,C\}, \{A,D,C\},\cdots\}$, and $|\mathcal{S}|=38$.
We set $\rho_1$ and $\rho_1$ as 0.2 and 0.8 respectively.
The formed clusters by the centralized clustering scheme are shown in Fig.~\ref{fig:final_clustering_LP}.
%$\{\{D,E,F\},\{A,C,G\},\{H,G\}\}$, the numbers of CCs are 2, 3, 3.
The resulted clustering solutions from SOC is shown in Fig.~\ref{fig:final_clustering_soc}.
We compare the average number of CCs achieved by different schemes, the results of ROSS\footnote{In this example network, both ROSS-DGA and ROSS-DFA and their size control variants form the same clusters)}, centralized and SOC are 2.66, 2.66, and 3 respectively. 
Note there is one singleton cluster $C(H)$ generated by SOC, which is not preferred.
When we only consider the clusters which are not singleton, the average number of CCs of SOC drops to 2.5. 

We investigate the schemes with four metrics.

\begin{itemize}
\item \textbf{The average number of CCs per non-singleton cluster.}
Non-singleton cluster refers the cluster whose cluster size is larger than 1.
%This metric denotes the robustness of the formed non-singleton clusters.
Comparing with the metric adopted by SOC~\cite{LIU_TMC11_2}, which is the average number of CCs of all the clusters, this metric provides a more accurate description of the robustness of the non-singleton clusters.
Having more CCs per non-singleton clusters means these clusters have longer life expectancy when the primary users' operation becomes more intense.
%which is biased as the singleton clusters (isolated CR users) don't get benefits from the cluster structure.
Although this metric doesn't disclose the information about the \textit{unclustered} CR nodes which are the synonyms of the singleton clusters, we still examine this metric as the number of CCs is involved in the utility adopted by all the variants of ROSS and SOC.

\item \textbf{Cluster sizes.}
%Specific clusters size is pursued in many applications due to energy preservation and the system design ~\cite{clustering_globecom11}.
We investigate the distribution of CRs residing in the formed clusters with different sizes.

\item \textbf{Robustness of the clusters against newly added PUs.}
We increase the number of PUs to challenge the non-singleton clusters, and count the number of the unclustered CR nodes.
This metric directly indicates the robustness of clusters from a more practical point of view, \ie as to the clusters formed for a given CRN and spectrum availability, how many CR nodes can still make use of the clusters when the spectrum availability decreases.
%robustness of clusters, as this metric directly shows how many nodes can make use of the cluster structure.
%When we investigate the performance with moderate and vigorous intensity of primary users' activities, 
%All the robust clustering schemes are proposed for a given CRN, when the available spectrum becomes scarce when the primary users' operation becomes more intense, the metric can illustrate the robustness of clusters by showing the unclustered CR nodes.
%This metric is closely related with cluster robustness, \ie the less CR nodes which are not  nodes means more CR nodes are included into non-singleton clusters which survive the primary users' influence.
%When we vary the intensity of primary users' activity, \eg from low to medium level by increasing the number of primary users, 
%The drawback of this metric is, it doesn't indicate the robustness of the non-singleton clusters when the primary users conquer more spectrum.

\item \textbf{Amount of control messages involved.}
We investigate the number of control messages involved in the clustering process.

\end{itemize}

Simulation consists of two parts, first we investigate the performance of centralized scheme and the distributed schemes in a small network, as there is no polynomial time solution available to solve the centralized problem.
In the second part, we investigate the performance of the proposed distributed schemes in the CRN with different scales and densities.
The following simulation settings is the same for both simulation parts.
CRs and PUs are deployed on a two-dimensional Euclidean plane.
%Complying with the system model, the CR node residing within another CR node's transmission range is seen as neighbour of that CR node.
The number of licensed channels is 10, each PU is operating on each channel with probability of 50\%.
CR users are assumed to be able to sense the existence of primary users and identify available channels.
All primary and CR users are assumed to be static during the process of clustering.
The simulation is written in C++, and the performance results are averaged over 50 randomly generated topologies, and the confidence interval corresponds to 95\% confidence level.

\subsection{Centralized Schemes vs. Decentralized Schemes}
There are 10 primary users and 20 CR users dropped randomly (with uniform distribution) within a square area of size $A^{2}$, where we set the transmission ranges of primary and CR users to $A/3$.
%There are 10 available channels. 
%With this setting, the average number of neighbours of one CR user is 4.8.
%Each primary user randomly occupies one channel, and 
When clustering scheme is executed, around 7 channels are available on each CR node.
The desired cluster size $\delta$ is 3.
As to the centralized scheme, the parameters used in the \textit{punishment} for choosing the clusters with undesired sizes are set as follows, $\rho_1 =  0.4$, $\rho_2 =  0.6$.

\subsubsection{Average number of CCs in Non-singleton Clusters}
\label{ccc_20}
\begin{figure*}[th]
\begin{multicols}{3}
    \includegraphics[width=\linewidth]{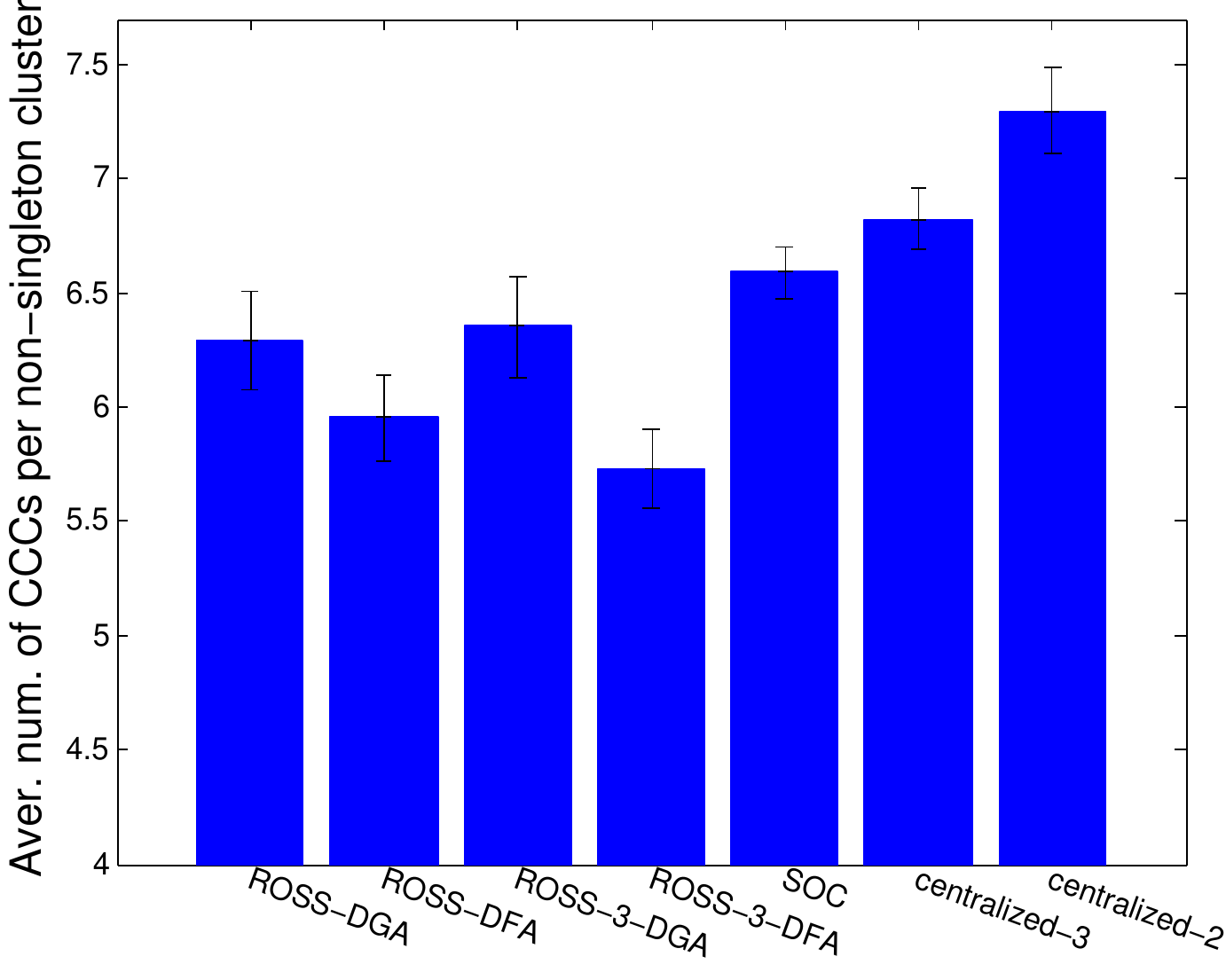}\par\caption{Number of common channels of non-singleton clusters}\label{ccc_per_nonsingleton}
    \includegraphics[width=\linewidth]{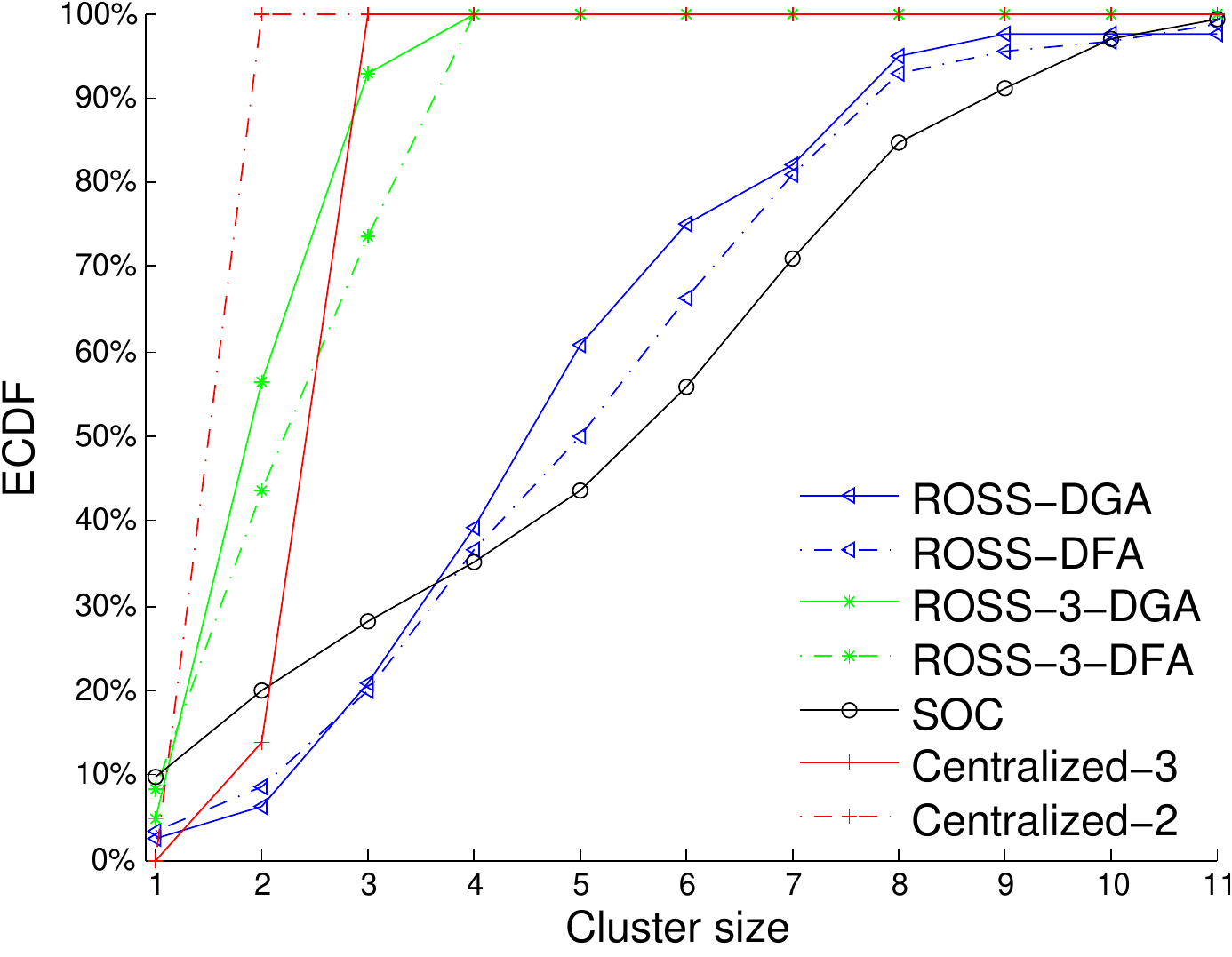}\par\caption{Cumulative distribution of CRs residing in clusters with different sizes}\label{size_control}    
    \includegraphics[width=\linewidth]{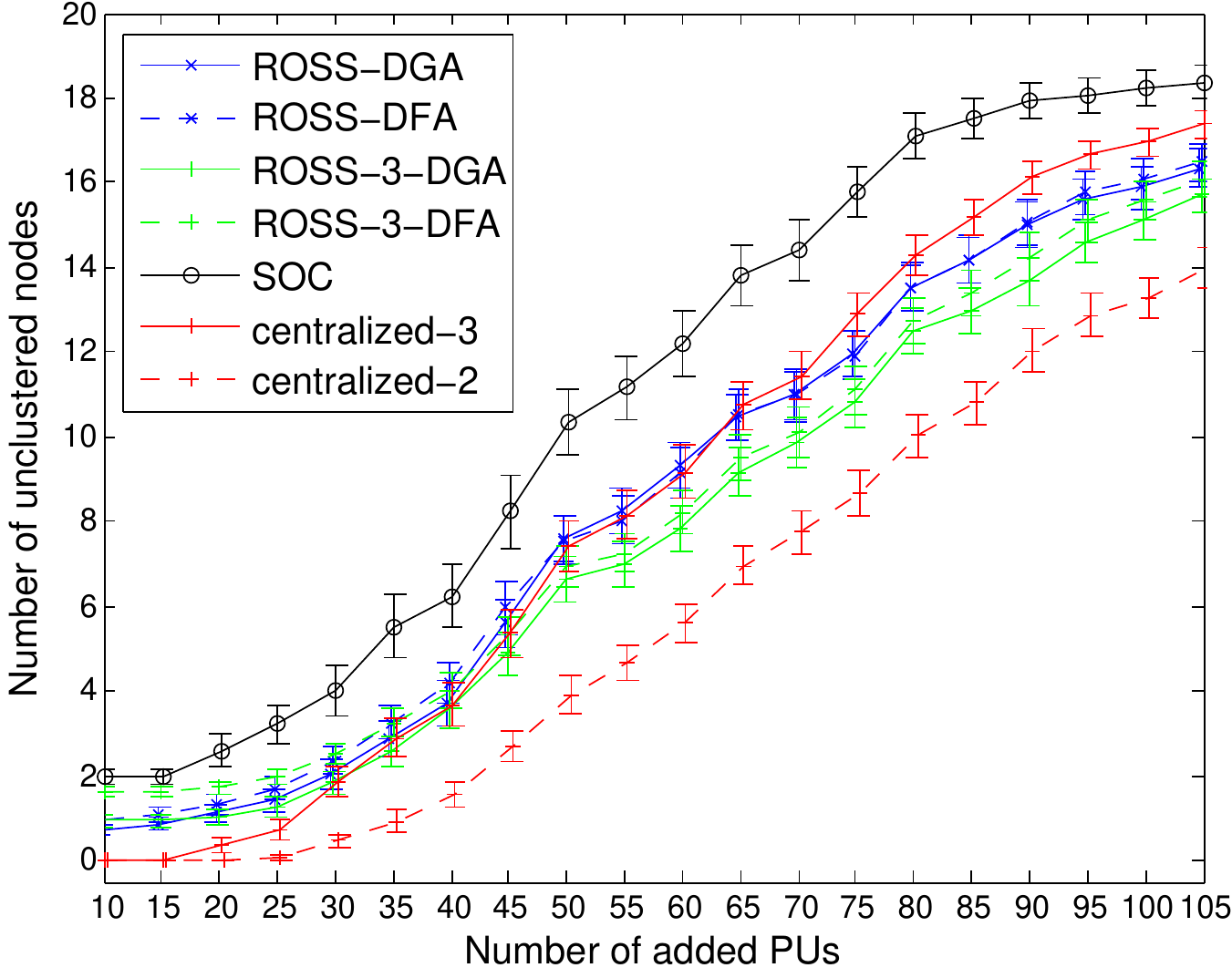}\par\caption{Number of unclustered CRs with decreasing spectrum availability}\label{singleton_clusters}
\end{multicols}
\caption{Comparison between the distributed and centralized clustering schemes ($N$ = 20)}
\label{compare_dis_centralized}
\end{figure*}

%We first have a look at the average number of common channels per non-singleton cluster.
% shows the average number of the CCs of non-singleton clusters,
%as the singleton clusters (in other words unclustered nodes) don't execute any functionalities of clusters, which are has be discussed in Section~\ref{intro}.
From Figure~\ref{ccc_per_nonsingleton}, we can see the centralized schemes outperform the distributed schemes.
Among the distributed schemes, SOC achieves the most CCs.
The reason is, SOC is liable to group the neighboring CRs which share the most abundant spectrum together, no matter how many of them are there, thus the number of CC of the formed clusters is higher.
%We have discussed the flaw of this metric as it doesn't convey the number of unclustered CR nodes, in fact, 
In the other hand, SOC generates the most unclustered CRs, which can be seen when we discuss the performance on the number of unclustered CR nodes. 
As to the variants of ROSS, we notice that the greedy mechanism increases CCs in non-singleton clusters significantly.
%, this is due to the procedure that debatable nodes greedily look for better affiliation to improve the number of CCs.
%We also notice that the size control feature doesn't affect the number of CCs for both ROSS-DGA and ROSS-DFA.
%Size control mechanism converts the large clusters into small ones, but meanwhile clusters with the desired sizes have to be made when forming smaller clusters is possible.
%, which can also be observed in the large scale CRN in Section~\ref{largeScaleCRN}.

%This is because the desired cluster size happens to be the average size of clusters generated by ROSS-DGA and ROSS-DFA, then the size control functionality doesn't play effect to increase the number of CCs.

\subsubsection{Cluster Size}
\label{cluster_size}
%\begin{figure}[h]
%  \centering
%  \includegraphics[width=0.8\linewidth]{cdf_clusterSize_20.pdf}
%  \caption{Cumulative distribution of CRs residing in clusters with different sizes.}
%  \label{size_control}
%\end{figure}

Figure~\ref{size_control} depicts the empirical cumulative distribution of the CRs in clusters of different sizes, from which we have two conclusions.
The first, SOC generates more unclustered CR nodes than other schemes.
The centralized schemes don't produce unclustered CR nodes in the simulation, the unclustered nodes generated by ROSS-DGA/DFA account for 3\% of the total CR nodes, as comparison, 10\% of nodes are unclustered when applying SOC.
ROSS-DGA and ROSS-DFA with size control feature generate 5\%-8\% unclustered CR nodes, which is due to the cluster pruning procedure (discussed in section~\ref{ross_p1_guarantee_ccc} and section~\ref{ross_p2_cluster_pruning}).
Second, the centralized schemes and cluster size control mechanism of ROSS generate clusters with the desired cluster size.
%When the desired size is 2, each generated cluster has two members, whereas when the desired size is 3, about 15\% CRs are formed into 2 node clusters.
As to ROSS-DFG and ROSS-DFA with size control feature, CR nodes reside averagely in clusters whose sizes are 2, 3 and 4.
%When ROSS-3-DFA is applied, most number of CRs are in 3 node clusters, nevertheless, slightly less nodes are found in 2 node and 4 node clusters, there are also considerable number of singleton clusters.
%ROSS-3-DGA decreases the clusters sizes and results in more 2 node clusters, the second most CRs are found in 3 node clusters.
The sizes of clusters resulted from ROSS-DGA and ROSS-DFA are disperse, but appear to be better than SOC, i.e., the 50\% percentiles for ROSS-DGA, ROSS-DFA and SOC are 4.5, 5, and 5.5, and the 90\% percentiles for the three schemes are 8, 8, and 9, the corresponding sizes of ROSS are closer to the desired size.
%Figure~\ref{size_control} shows distributed clustering schemes are not able to control cluster sizes perfectly, but ROSS-DGA and ROSS-DFA eliminate the clusters whose size diverges largely with the desired one, \ie single node clusters and clusters with size of 13 and 14.

\subsubsection{Robustness of the clusters against newly added PUs}
%When the number of PUs in CRN increases, or their operation becomes more intensive, some clusters don't seize any CCs any more, so that the cluster members and the cluster heads become unclustered, or singleton clusters.
%If there is no common channels available any more because of the new added PRs, the cluster is regarded as destroyed and the former cluster member CRs become unclustered CRs or in other words singleton clusters.
%We investigate the number of singleton clusters with primary users whose intensity of activities are varying.
%We will obtain two observations by examining this metric.
%First, we will know how many CR nodes are unclusterred by applying the robust clustering schemes in a given CRN.
In this part of simulation, we put PUs sequentially into CRN to decrease the available spectrum.
10 PUs are in the network in the beginning, then extra 19 batches of PUs are added sequentially, where each batch includes 5 PUs. 
%The transmission range and channel occupancy of the new PU is the same with the previous ones, \ie transmission range is $A/3$, and one channel out of 10 is randomly chosen to operate.

Figure~\ref{singleton_clusters} shows certain clusters can not maintain and the number of unclustered CR nodes grows when the number of PUs increases.
The centralized scheme with desired size of 2 generates the most robust clusters, meanwhile, SOC results in the most vulnerable clusters.
The centralized scheme with desired size of 3 doesn't outperform the variants of ROSS, because pursuing cluster size prevents forming the the clusters with more CCs.
In contrary, the variants of ROSS generate some smaller clusters which are more likely to maintain when there are more PUs.
%The comparison on cluster sizes will be given in Section~\ref{cluster_size}.
% when the number of PRs is 10$\sim$30, when number of PUs is 30$\sim$60, same amount of unclustered CRs are generated with variants of ROSS.
%When there are 75 and more new PRs, centralized scheme with cluster size of 3 results in more unclustered CR nodes than variants of ROSS.
%ROSS with size control distinguishes itself as the size control avoid the appearance of the clusters with large size.

%As to the variants of ROSS, the greedy process in ROSS improves the performance. 

%Greedy strategy adopted in the second phase of ROSS improves the robustness of clusters, i.e. ROSS-DGA exceeds ROSS-DFA, and ROSS-$\delta$-DGA surpasses ROSS-$\delta$-DFA.
%When the debatable CRs greedily update their affiliations, one of the metrics is the maximum increase of CCs of the demanding clusters. 
%This observation complies with the result shown in Figure~\ref{ccc_per_nonsingleton}.
%Meanwhile, sizes of more clusters become smaller also contributes more robustness.

%\begin{figure}[h!]
%  \centering
%  \includegraphics[width=0.8\linewidth]{ccc_20.pdf}
%  \caption{Number of common channels for non-singleton clusters, the numbers in the names of schemes annotate the desired cluster size.}
%  \label{ccc_per_nonsingleton}
%\end{figure}
%
%
%\begin{figure}[h]
%  \centering
%  \includegraphics[width=0.8\linewidth]{survival_rate_20.pdf}
%  \caption{Number of CRs which are not included in any clusters}
%  \label{singleton_clusters}
%\end{figure}

\subsubsection{Control Signaling Overhead}
%Different from the clustering schemes proposed in ~\cite{LIU_TMC11_2, clustering_globecom11}, 
%There are two phases for any variants of ROSS.
%Clusters are formed in the first phase, in the second phase, cluster membership is decided so that each node only resides in one cluster.
%Control message exchanges between CR nodes are involved in both phases.

In this section we compare the overhead of signaling involved in different clustering schemes.
%In order to highlight the amount of control signalling only for clustering, 
We don't consider the the control messages which are involved in neighborhood discovery, which is the premise and deemed to be the same for all clustering schemes.
According to~\cite{complexity_aggregation_2011}, the message complexity is defined as the number of messages used by all nodes.
To have the same metric to compare, we count \textit{the number of transmissions of control messages}, without distinguishing broadcast or uni-cast control messages.
This metric is synonymous with \textit{the number of updates} discussed in Section~\ref{ross}.

As to ROSS, the control messages are generated in both phases.
In the first phase, when a CR node decides itself to be the cluster head, it broadcasts a message containing its ID, cluster members and the set of CCs in its cluster.
In the second phase, a debatable node broadcasts its affiliation to inform its claiming clusters, then the cluster heads of the claiming clusters broadcast message about the new cluster members if they are changed due to the debatable node's decision.
The upper bound of the total number of the control messages involved in cluster formation is analyzed in Theorem~\ref{clustering:theorem} and Section~\ref{clustering:phaseII:game}.

The comparison scheme SOC involves three rounds of execution. 
In the first two rounds, every CR node maintains its own cluster and seeks either to integrate neighboring clusters or to join one neighboring cluster.
The final clusters are obtained in the third round. 
In each round, every CR node is involved in comparisons and cluster mergers.
%Comparing with the second phase of ROSS, only debatable nodes to communicate with cluster heads to clarify their membership.
%The signalling overhead for centralized scheme comes from two processes, the the process of collecting information to the centralized controller, and the process that the controller spreads the clustering result to all the CR nodes.

The centralized scheme is conducted at the centralized control device, but it involves two phases of control message transmission.
The first phase is information aggregation, in which every CR node's channel availability and neighborhood is transmitted to the centralized controller.
iIn the second phase, the control broadcasts the clustering solution, which is disseminated to every CR node.
We adopt the algorithm proposed in~\cite{Efficient_broadcasting_gathering_adhoc} to broadcast and gather information as the algorithm is simple and self-stabilizing.
This scheme needs building a backbone structure to support the communication. 
We apply ROSS to generate cluster heads which serve as the backbone, and the debatable nodes are used as the gateway nodes between the backbone nodes.
As the backbone is built for one time and supports the transmission of control messages later on, we don't take account the messages involved in building the backbone.
As to the process of information gathering, we assume that every cluster member sends the spectrum availability and its ID to its cluster head, which further forwards the message to the controller, then the number of transmissions is $N$.
As to the process of dissemination, in an extreme situation where all the gateway and the backbone nodes broadcast, the number of transmissions is $h + m$, where $h$ is the number of cluster heads and $m$ is number of debatable nodes.
%, $d$ is the average number of demanding clusters for each debatable node.

\begin{figure}[ht!]
  \centering
  \includegraphics[width=0.6\linewidth]{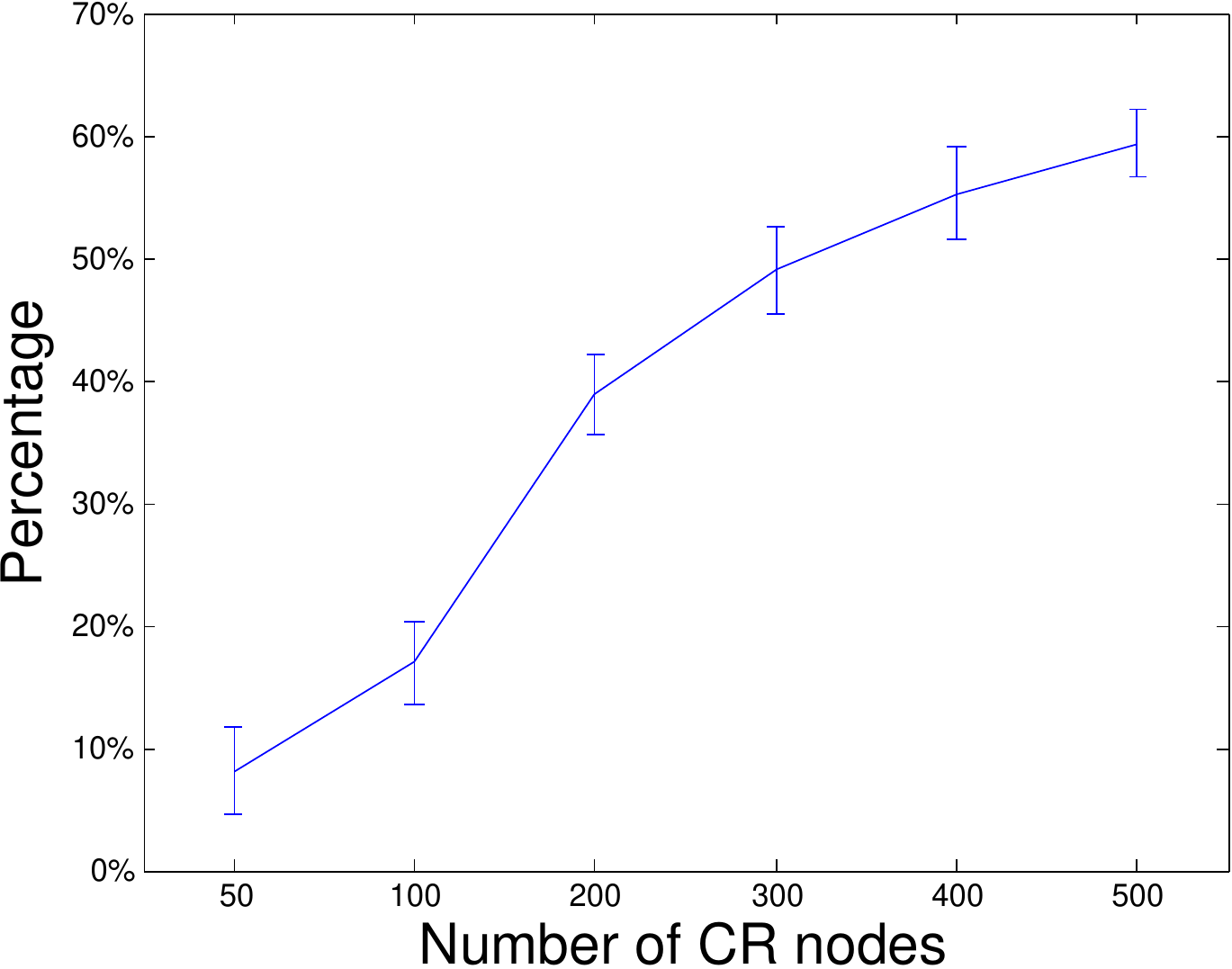}
  \caption{The percentage of debatable nodes after phase I of ROSS.}\label{percentage_overlapping_node}
\end{figure}

\begin{figure}[ht!]
  \centering
  \includegraphics[width=0.6\linewidth]{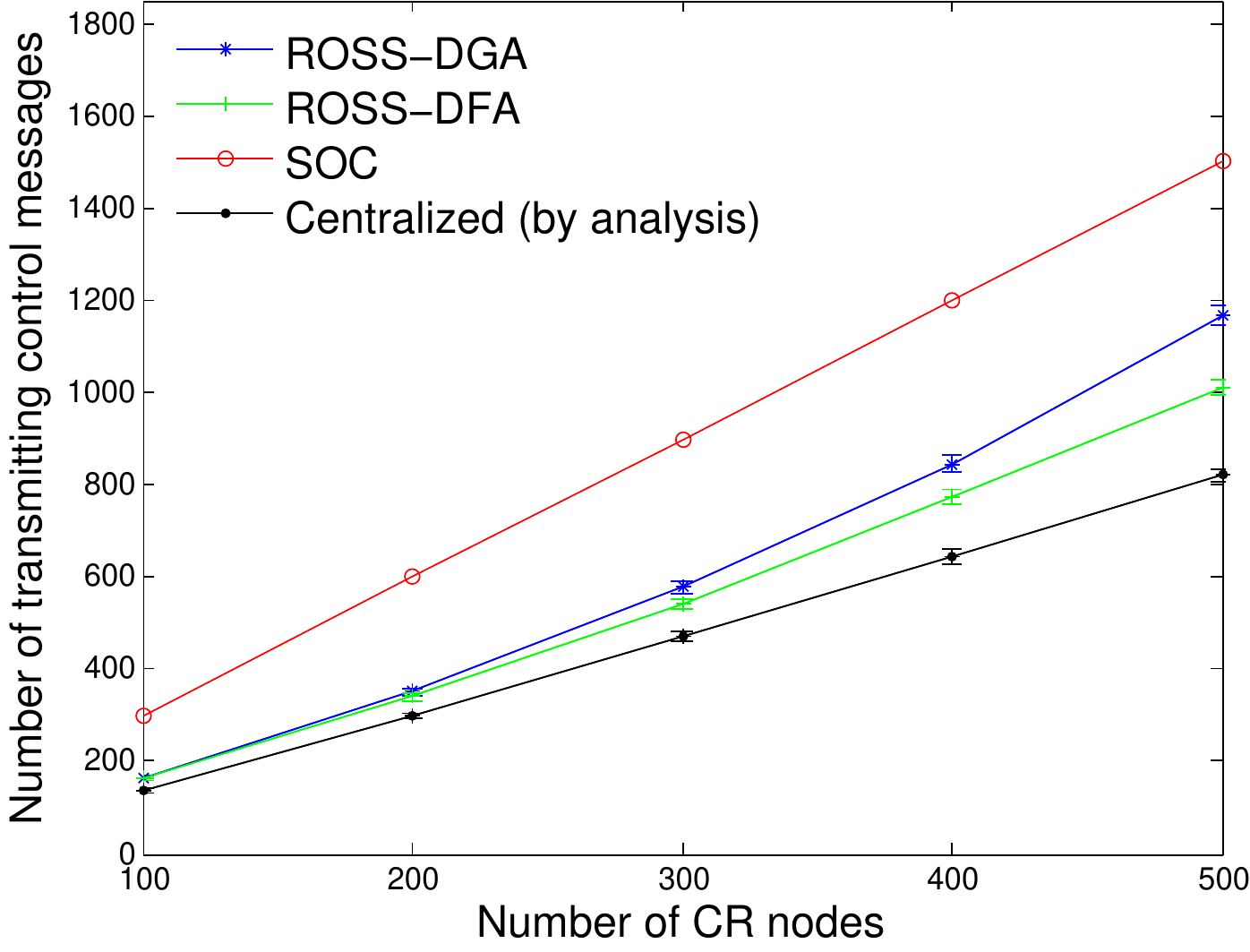}
  \caption{Quantitative amount of control messages.}
  \label{control_msg}
\end{figure}

The number of control messages which are involved in ROSS variants and the centralized scheme is related with the number of debatable nodes.
Figure~\ref{percentage_overlapping_node} shows the percentage of debatable nodes with different network densities, from which we can obtain the value of $m$.
%
%As we adopt a simplified communication model, node's transmission is not influenced by collision or interference, 
%Our simulation doesn't consider the behaviour in the physical layer, 
%
%
%Assume we use OSPF~\cite{BCJ10} to aggregate and disseminate information, then the best and worst complexity for is $\mathcal{O}(E)$, where $E$ is the number of edges in the graph which corresponds to the network.
%The minimum number of edges is $n-1$ when the nodes form a line and each node has at more two neighbours, and the maximum number is $N*(N-1)/2$ when the nodes form a complete graph.
%Thus the best message complexity of the centralized scheme is $\mathcal{O}{(N)}$ and the worst is $\mathcal{O}{(N^2)}$.
%
%1. update membership to form X1, 
%2. broadcast new X1, form new X2
%3. broadcast X3
%The complexity parameters are the number of nodes $n$ in network, number of clusters $h$.
Table \ref{tab_overhead} shows the message complexity, quantitative amount of the control messages, and the size of control messages.
Figure~\ref{control_msg} shows the analytical result of the amount of transmissions involved in different schemes.
%the upper bound of the number of transmissions of ROSS, and the analytic number of transmissions of the centralized scheme. 

%\begin{table}[hc]
%\centering
%\caption{Singalling overhead. Notations: $n$-number of CR nodes in CRN, $h$-number of cluster heads, $m$-number of debatable nodes, $c$-number of demanding clusters.}\label{tab_overhead}
%\begin{tabular}{|p{2.2 cm}|p{1.5 cm}|p{3.7 cm}|}
%\hline
% Scheme 					&   Number of transmissions 					& Content of message \\ \hline
% ROSS-DGA, ROSS-$\delta$-DGA 		&   $h+2*m^2c$ (upper bound)
%				& $ID_{H_C}$ and $K_C$ for $h+m^2c$ times, notification to join one cluster for $m^2c$ times					\\ \hline
% ROSS-DFA, ROSS-$\delta$-DFA 		&   $h+ 2m$	 (upper bound) 						& $ID_{H_C}$ and $K_C$ for $h+m$ times, notification to join one cluster for $m$ times	 					\\ \hline
% SOC 						&   $3*n$					& $\{K_i\}, i\in M\subseteq \texttt{Nb}_i$						\\ \hline
% Centralized				&	$n$						& $\{C\}$         	\\ 
% \hline
%\end{tabular}
%\end{table}

\begin{center}
\begin{table*}[!htb]
\caption{Signalling overhead}\label{tab_overhead}
{\renewcommand{\arraystretch}{1.15} %<- modify value to suit your needs
{\small
\hfill{}
\begin{threeparttable}
\begin{tabular}{|C{2 cm}|C{2.8 cm}|C{3.15 cm}|C{7.7 cm}|}
\hline
 Scheme 				&Message Complexity 	&   Quantitative number of messages 		& Content of message (size of message) 									\\ \hline
 ROSS-DGA, ROSS-$\delta$-DGA 	&$\mathcal{O}(N^3)$ (worst case)		&   $h+2m^2d$ (upper bound)  				&   \multirow{2}{*}{\parbox{7.7cm}{Cluster head $i$ broadcasts channel availability on all cluster members ($|C(i)| |\mathcal{K}|$ bytes); Cluster member $i$ broadcasts the new individual connectivity $d_i$ after being included in one or more clusters (1 byte)}}								\\ \cline{1-3}
 ROSS-DFA, ROSS-$\delta$-DFA 	&$\mathcal{O}(N)$ (worst case)		&   $h + 2m$	 (upper bound) 					& 	      												\\ \hline
 SOC 					&$\mathcal{O}(N)$		&   $3N$									& Every CR node $i$ broadcasts channel availability on all cluster members ($|C(i)| |\mathcal{K}|$ bytes)
 \\ \hline
 Centralized			&$\mathcal{O}(N)$			&	$h + m + N$ (upper bound)~\cite{Efficient_broadcasting_gathering_adhoc}		& clustering result (2$N$ bytes) \tnotex{tnote:robots-r1} 					\\ \hline
\end{tabular}
    \begin{tablenotes}
      \item\label{tnote:robots-r1}Assuming the data structure of the clustering result is in the form of \{ Node ID $i$, cluster head ID h($C$) where $i\in C$, for every $i\in \mathcal{N}$ \}.
    \end{tablenotes}
    \end{threeparttable}
}
}
\hfill{}
\end{table*}
\end{center}

\subsection{Comparison among the Distributed Schemes}
\label{largeScaleCRN}
In this section we investigate the performances of distributed clustering schemes in CRN with different network scales and densities.
The transmission range of CR is $A/5$, PU's transmission range is $2A/5$.
The initial number of PU is 30.
%The number of CR is 100, 200 and 300, and the average number of neighbours of each CR is 9.5, 20, and 31.
The desired sizes adopted are listed in the Table~\ref{Simulation_para}, which is about 60\% of the average number of neighbours.
When run ROSS, the parameter $t$ which is used to control cluster size in phase I is 1.3.

\begin{table}[ht]
\caption{}
\label{Simulation_para}
{\small
\hfill{}
\begin{tabular}{|L{3.7 cm}|C{1 cm}|C{1 cm}|C{1 cm}|}
\hline
Number of CRs			& 100 	&  200 					& 300 \\ \hline
Average num. of neighbours 	&9.5	&   20		& 31  \\ \hline
Desired size $\delta$ 	& 6	&   12 						& 20      \\ \hline
\end{tabular}
}
\hfill{}
\end{table}

\subsubsection{Number of CCs per Non-singleton Clusters}

\begin{figure}[ht!]
  \centering
  \includegraphics[width=.7\linewidth]{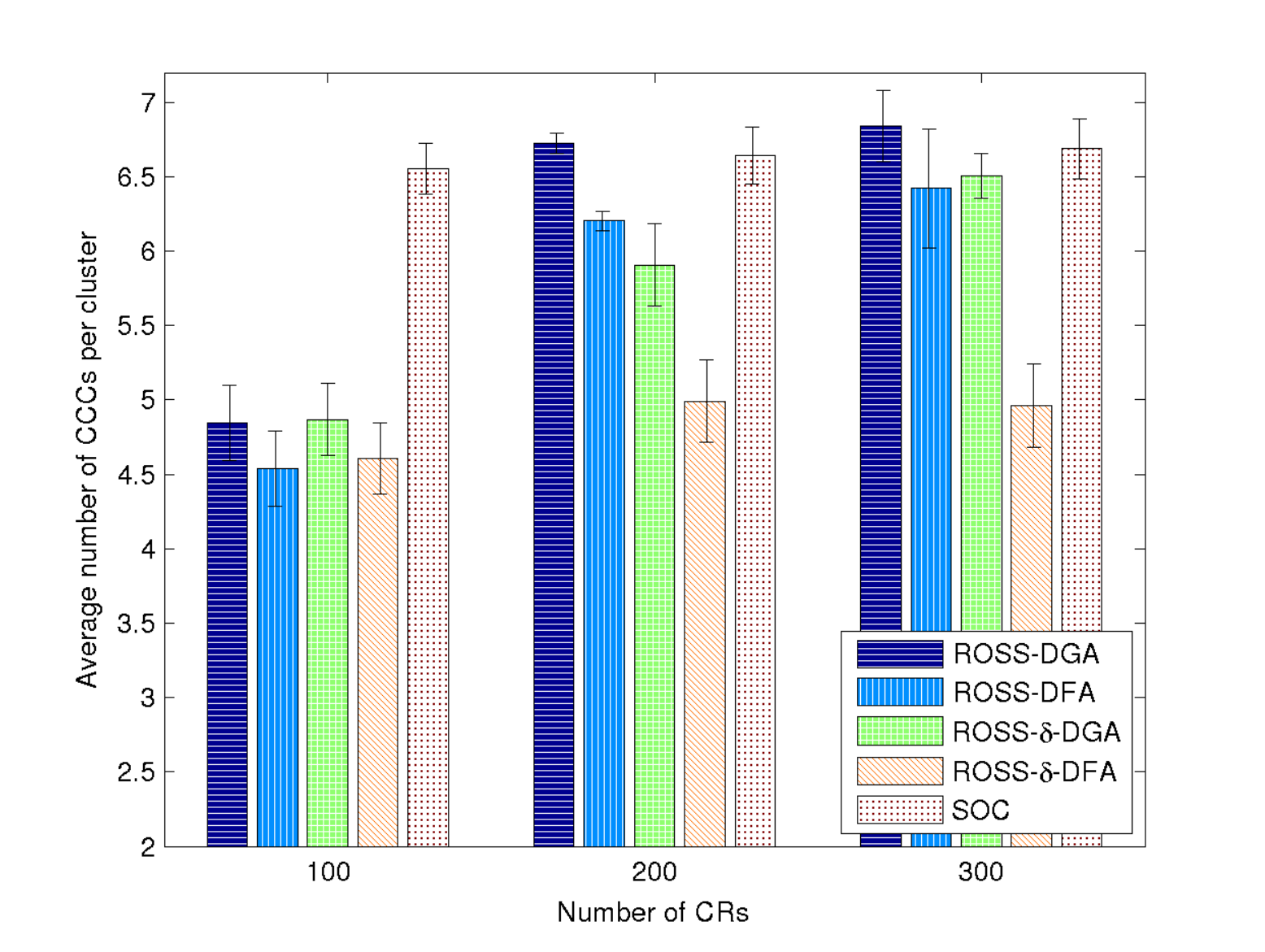}
  \caption{Number of common channels of non-singleton clusters.}
  \label{ccc_large_scale}
\end{figure}

Figure~\ref{ccc_large_scale} shows the average number of CCs of the non-singleton clusters.
We notice that SOC achieves the most CCs per non-singleton cluster, although the lead over the variants of ROSS shrinks significantly when $N$ increases.
%\ie when $N=300$, the number of CCs achieved by ROSS variants (except for ROSS-$\delta$-DFA) is almost the same as that resulted from SOC.
%This means SOC out performs in terms of the average number of CCs per non-singleton cluster when network is sparse.
%This is also observed in the evaluation in Section \ref{ccc_20} where $N=20$.
%When the network becomes denser, even this metric favours SOC as discussed in the beginning of Section~\ref{performance}, ROSS-DGA achieves even more CCs than SOC, and ROSS-DFA and ROSS-$\delta$-DGA increase the number of CCs visibly.

\subsubsection{Robustness of the clusters against newly added PUs}
\begin{figure*}[t]
\begin{multicols}{3}
    \includegraphics[width=\linewidth]{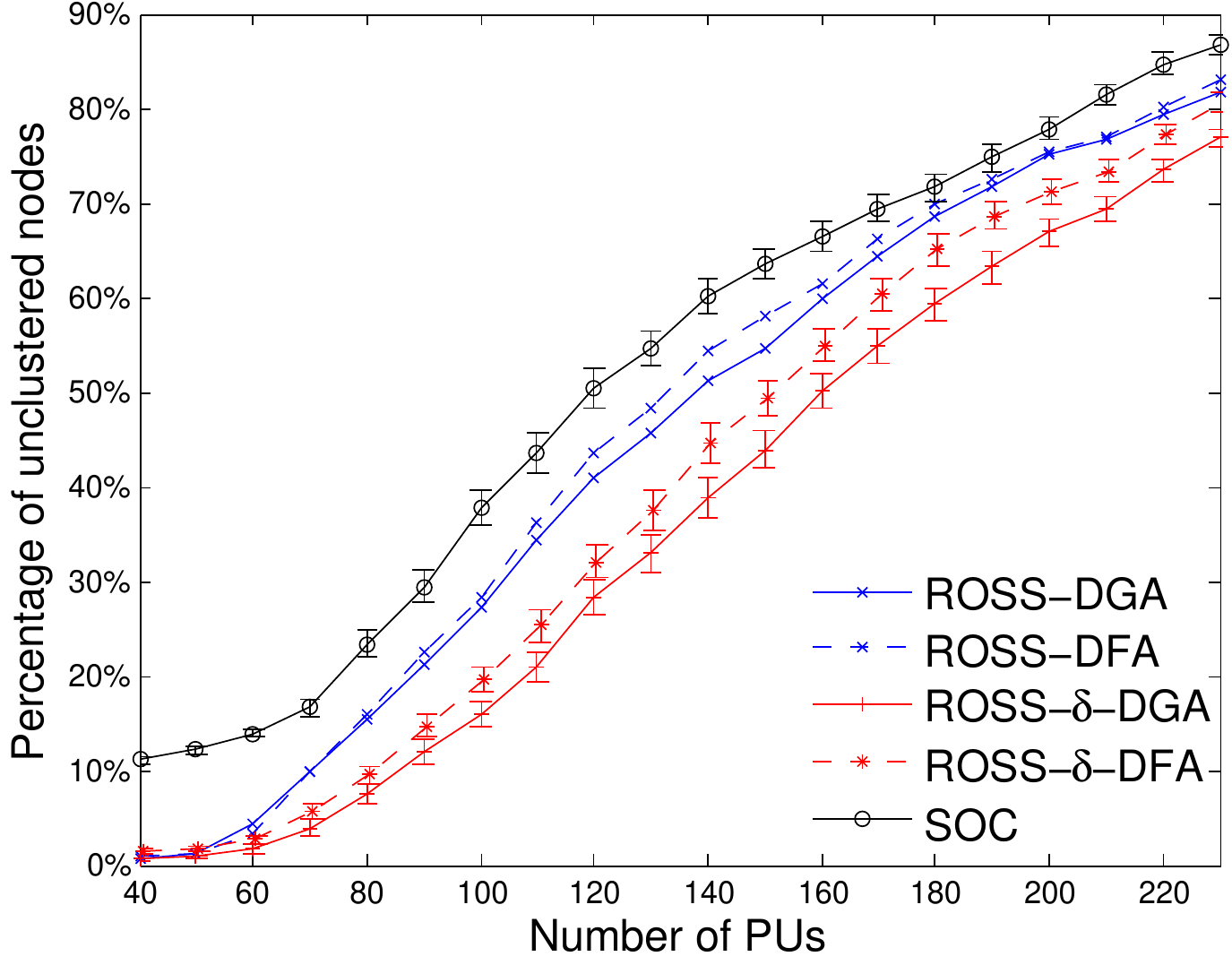}\par\caption{100 CRs}\label{singleton_clusters_100}
    \includegraphics[width=\linewidth]{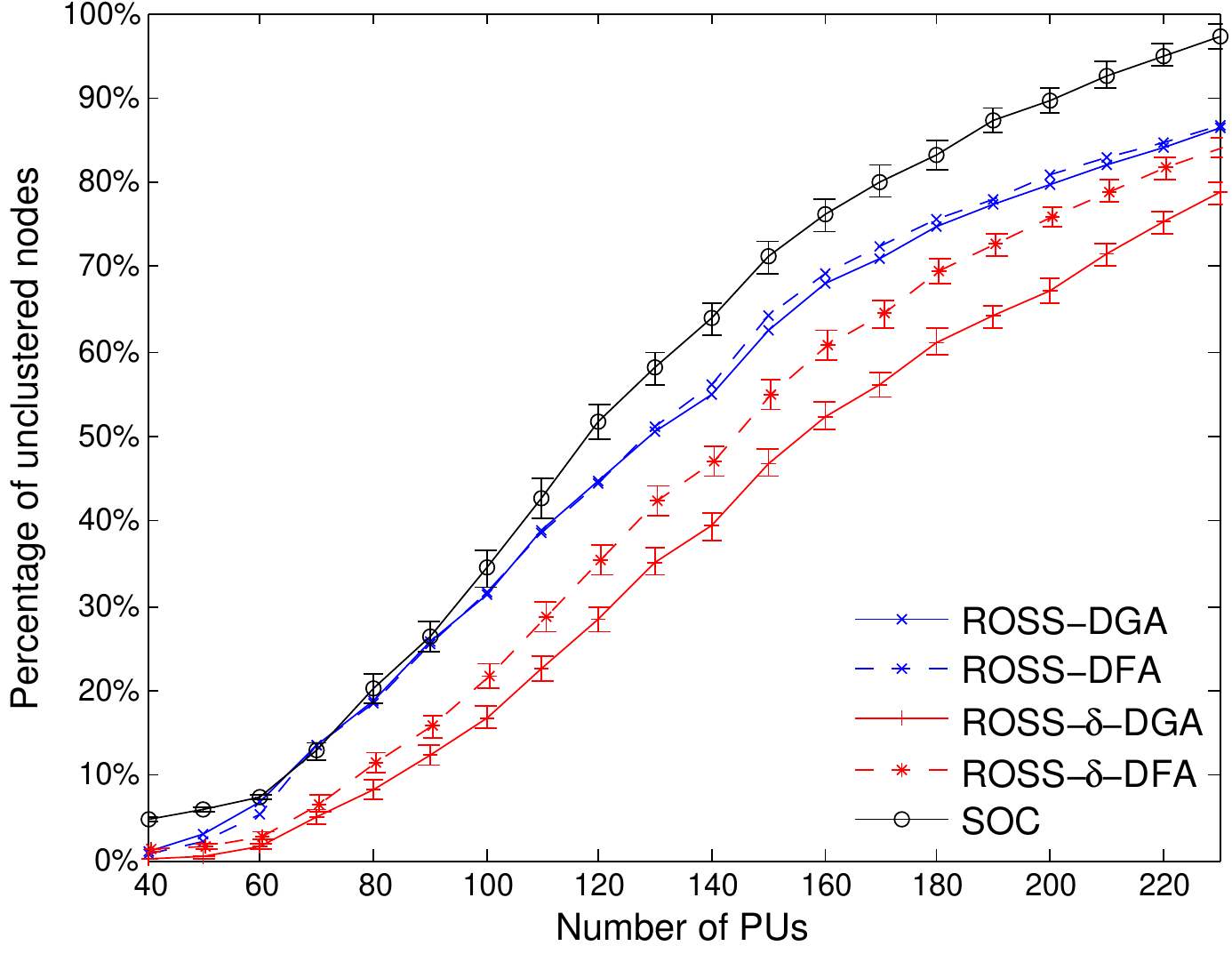}\par\caption{200 CRs}\label{singleton_clusters_200}
    \includegraphics[width=\linewidth]{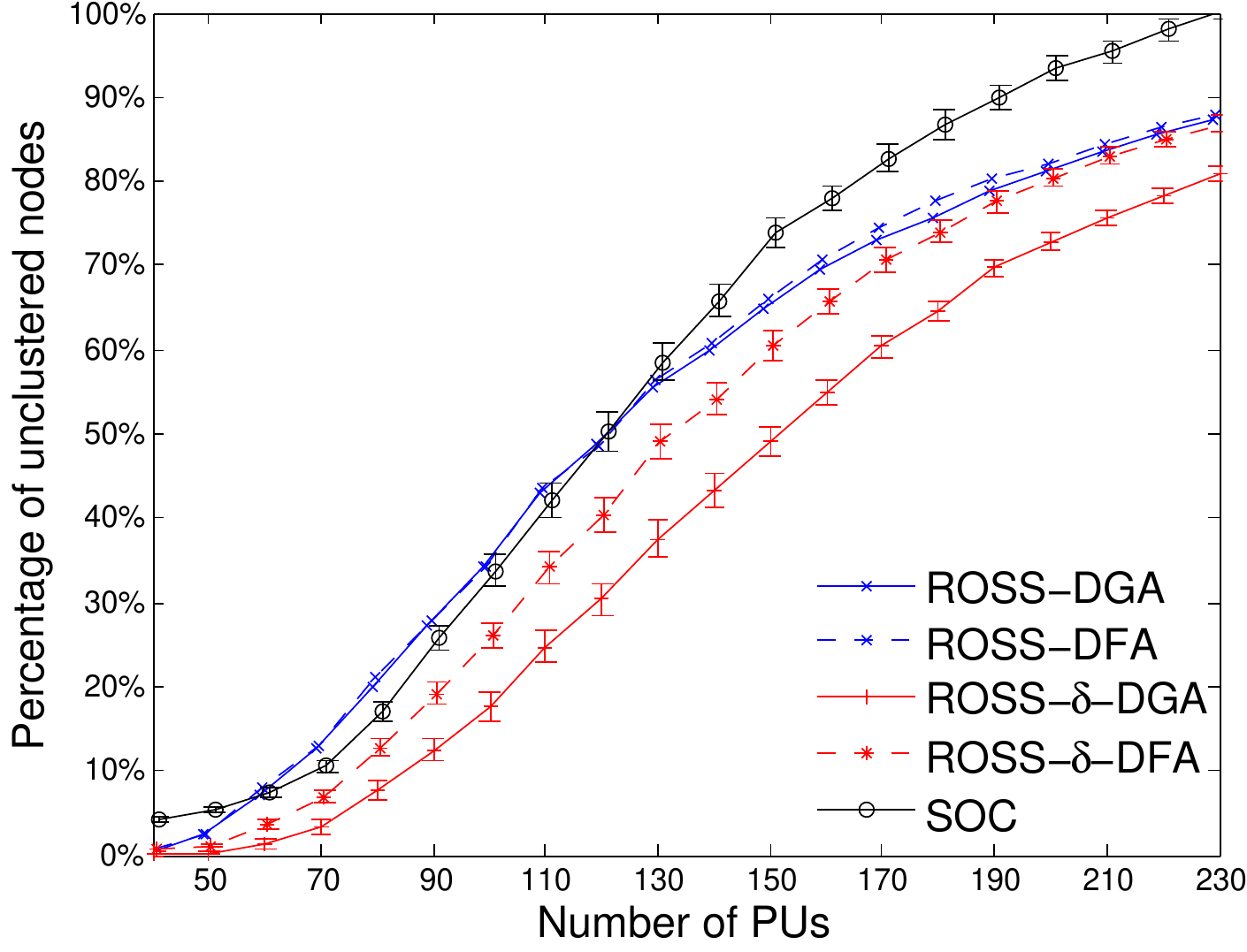}\par\caption{300 CRs}\label{singleton_clusters_300}
\end{multicols}\caption{Percentage of CR nodes which are not included in any non-singleton clusters}
\label{unclustered_100_200_300}
\end{figure*}

We add extra 20 batches of PUs sequentially in the CRN, where each batch includes 10 PUs. 
Figure~\ref{singleton_clusters_100} and \ref{singleton_clusters_200} show that when $N=100$ and 200, more unclustered CR nodes appear in the CRN where SOC is applied.
When the network becomes denser, as shown in Figure~\ref{singleton_clusters_300}, ROSS-DGA/DFA generate slightly more unclustered CR nodes than SOC when new PUs are not many, but SOC's performance deteriorates quickly when the number of PUs becomes larger.
We only show the average values of the variants of ROSS as their confidence intervals overlap.
When applying ROSS with size control mechanism, significantly less unclustered CR nodes are generated.
Besides, the greedy mechanism moderately strengthens the robustness of the clusters.

%When the network is denser, the improvement on cluster sizes and robustness by the greedy search in the membership clarification phase is more obvious.

\begin{figure*}[t]
\begin{multicols}{3}
    \includegraphics[width=\linewidth]{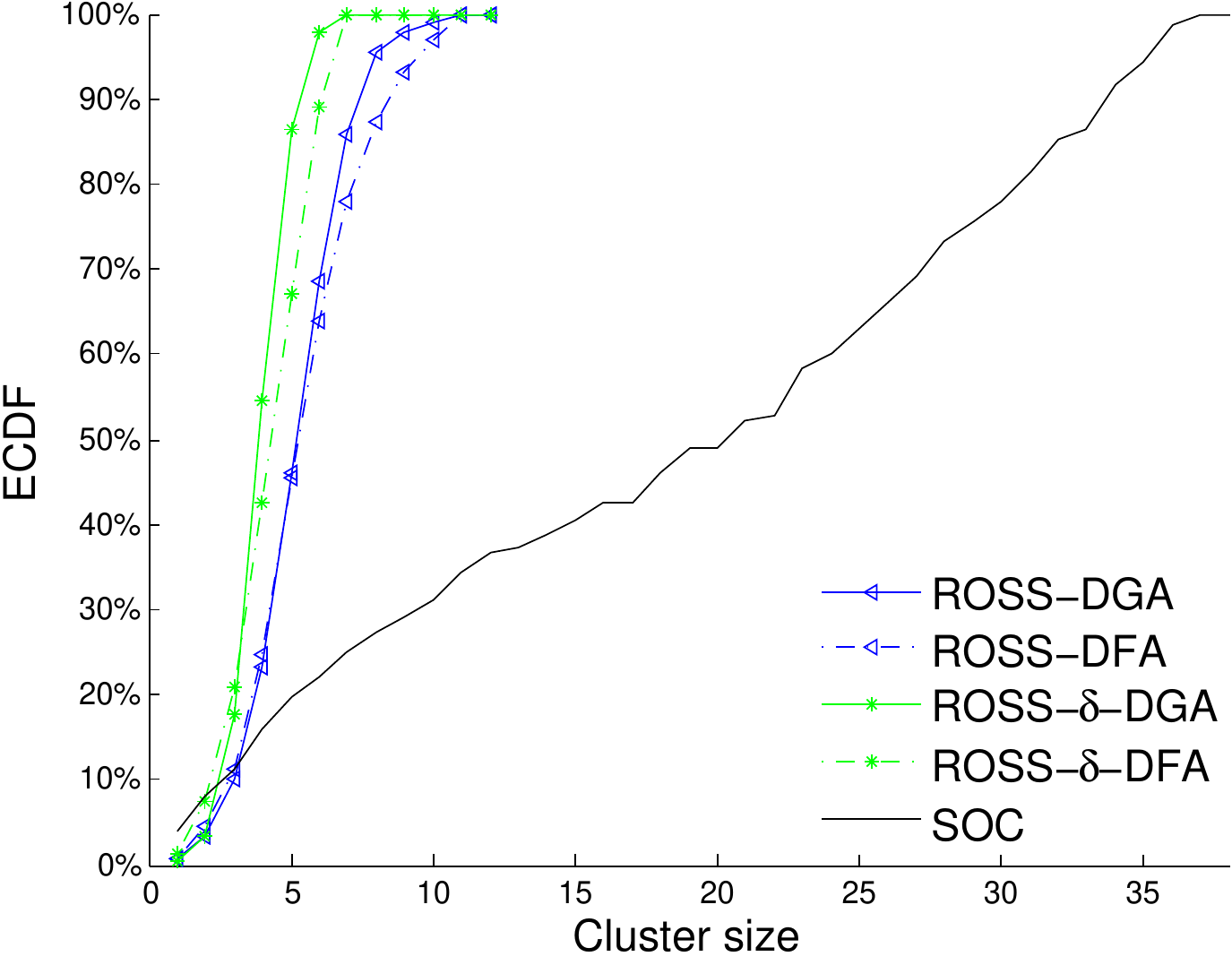}\par\caption{100 CRs, 30 PUs in network}\label{cdf_clusterSize_100}
    \includegraphics[width=\linewidth]{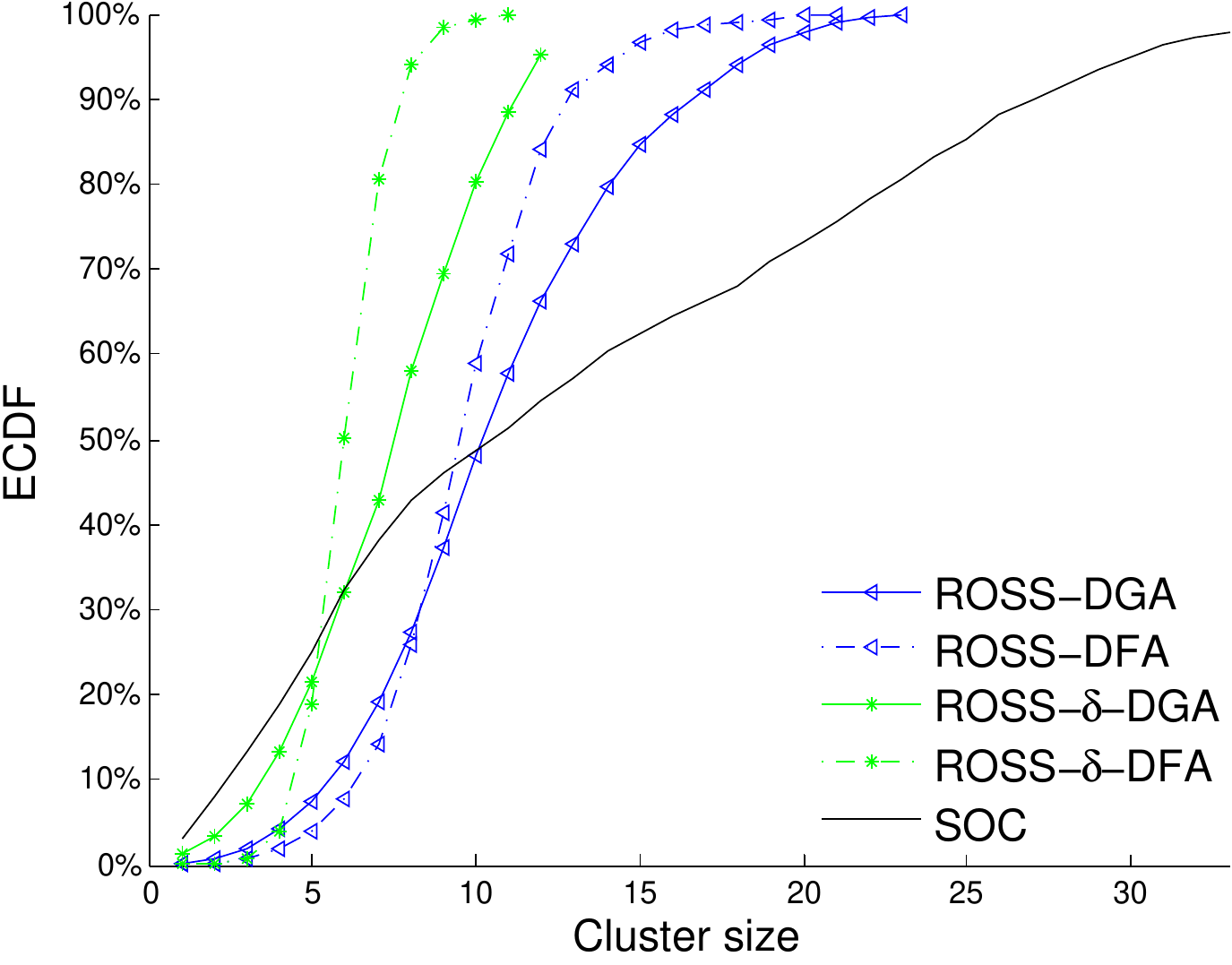}\par\caption{200 CRs, 30 PUs in network}\label{cdf_clusterSize_200}
    \includegraphics[width=\linewidth]{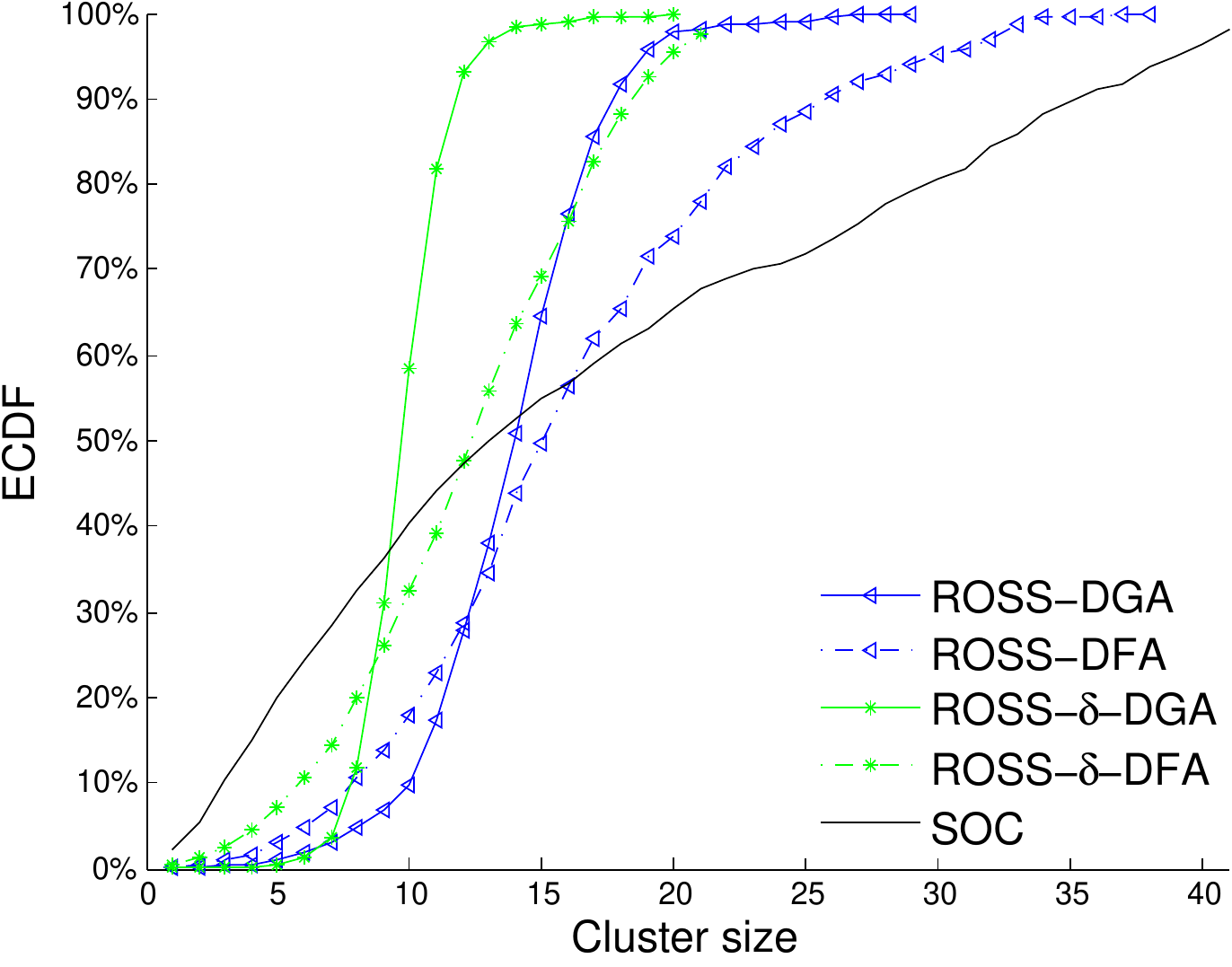}\par\caption{300 CRs, 30 PUs in network}\label{cdf_clusterSize_300}
\end{multicols}
\caption{Cumulative distribution of CRs residing in clusters with different sizes}
\label{cdf_100_200_300}
\end{figure*}

\subsubsection{Cluster Size Control}
\begin{figure}[!h]
  \centering
   \includegraphics[width=0.7\linewidth]{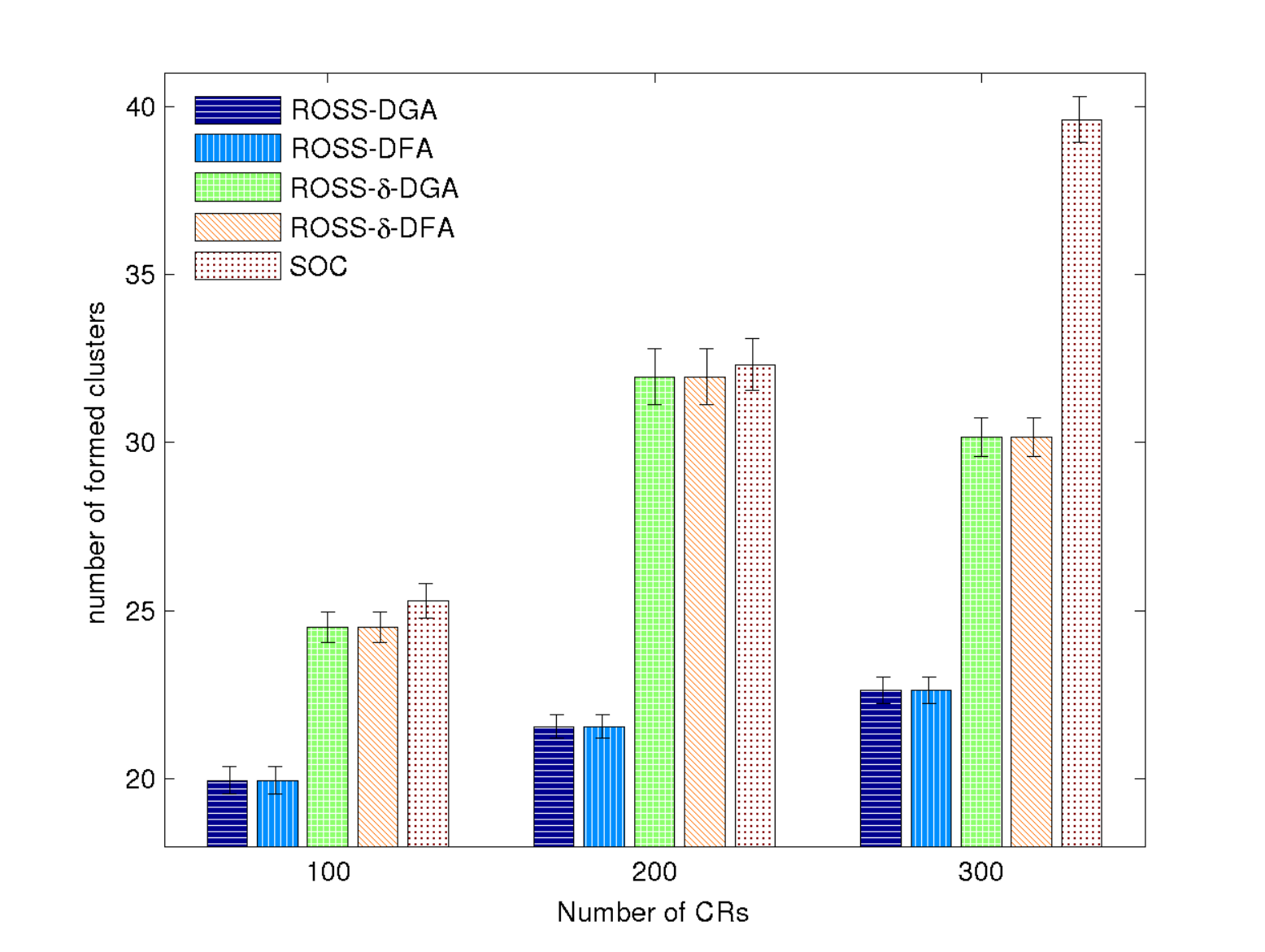}
  \caption{The number of formed clusters.}
  %, there are $x=6$ when $N=100$, $x=12$ when $N=200$, $x=21$ when $N=300$, which is around $2/3$ of the number of average neighbours.
  \label{nClusters_largeNetwork}
\end{figure}

Figure~\ref{nClusters_largeNetwork} shows when the network density scales up, the number of formed clusters by ROSS increases by smaller margin, and that generated by SOC increases linearly.
This result coincides with the analysis in Section~\ref{ross_p2_cluster_pruning}.
%When the network becomes denser, more clusters are generated by SOC compared with ROSS variants.
To better understand the distribution of the sizes of formed clusters, we depict the empirical cumulative distribution of CR nodes in clusters with different sizes in Figures~\ref{cdf_clusterSize_100}~\ref{cdf_clusterSize_200}~\ref{cdf_clusterSize_300}.

The sizes of clusters generated by ROSS-DGA and ROSS-DFA span a wider range than ROSS with size control feature.
Most of the generated clusters are smaller than the average number of neighbours, which is roughly equal with the 95\% percentile of the ROSS-DGA curve.
The 50\% percentile of the ROSS-DGA curve is roughly the desired size $\delta$.
When the variants of ROSS with size control feature are applied, the sizes of the most generated clusters are smaller than $\delta$.
As to the curves of SOC, the 95\% percentiles are 36, 30, and 40 in respective networks.
From Figure~\ref{cdf_100_200_300}, we conclude that the sizes of the clusters generated by ROSS are limited by the network density, the sizes of the clusters formed by ROSS with size control feature are restricted by the desired size.
In contrary, the clusters generated from SOC demonstrate strong divergence on cluster sizes.

\subsection{Insights Obtained from the Simulation}
The centralized clustering scheme is able to form the clusters which satisfy the requirement on cluster size strictly, and the clusters are robust against the PUs' activity, besides, it generates the smallest control overhead in the process of clustering.

As distributed schemes, the variants of ROSS outperform SOC considerably on three metrics.
The variants of ROSS generate much less singleton clusters than SOC, and the resulted clusters are robuster than SOC when facing the newly added PUs.
The signaling overhead involved in ROSS is about half of that needed for SOC, and the signaling messages are much shorter that the latter.
The sizes of the clusters generated by ROSS demonstrate smaller discrepancy than that of SOC.
Besides, the ROSS variants with size control features achieve similar performance to the centralized scheme in terms of cluster size, and the cluster robustness is similar when applying the variants of ROSS and the centralized scheme respectively.

As to the variants of ROSS, the greedy mechanism in ROSS-DGA helps to improve the performance on cluster size and cluster robustness at the cost of mildly increased signaling overhead.
We also notice that  as a metric, the number of CCs per non-singleton cluster doesn't indicate the robustness of clusters as shown in Figure~\ref{singleton_clusters} and \ref{unclustered_100_200_300}, although it is adopted as the metric in the formation of clusters.

\section{Conclusion}
\label{conclusion}
In this paper we investigate the robust clustering problem in CRN extensively.
%generate robust clusters against primary users' influence, so that more CR users can benefit from operative decision making provided by the clusters.
We provide the mathematical description of the problem and prove the NP hardness of it.
We propose both centralized and distributed schemes ROSS, the cluster structure generated by them has longer time expectancy against the primary users' activity.
Besides, the proposed schemes can generate clusters with desired sizes.
The congestion game model in game theory is used to design the distributed schemes.
%A Light weighted clustering scheme ROSS is also proposed.
Through simulation and theoretical analysis, we find that distributed schemes achieve similar performance with centralized optimization in terms of cluster robustness, signaling overhead and cluster sizes, and outperform the comparison distributed scheme on the above mentioned metrics.
%, which is important to form clusters which maintains unbroken to the greatest extent possible under primary users' activity.
%These schemes outperform other distributed clustering scheme in terms of both cluster survival ratio and control overhead.

%The clusters resulted from ROSS-DGA and its faster version ROSS-DFA are less vulnerable compared with other distributed clustering schemes, and demonstrates similar survival rate with centralized scheme under primary users' influence.
%An light weighted cluster size control mechanism is contained in both ROSS-DGA and ROSS-DFA, which is advantageous for cooperative sensing and network operation with clusters.
%Furthermore, considerable less control messages are generated when compared with other clustering schemes.

The shortcoming of distributed scheme ROSS is it doesn't generate clusters whose sizes exceed the cluster head's neighborhood.
The reason is with ROSS, cluster heads form clusters on the basis of their neighborhood, and don't involve the nodes which are outside the neighborhood.
In the other way around, forming big cluster which extends a cluster head's neighborhood has limited application scenarios, as multiple hop communication and coordination are required within these clusters.

% if have a single appendix:
%\appendix[Proof of the Zonklar Equations]
% or
%\appendix  % for no appendix heading
% do not use \section anymore after \appendix, only \section*
% is possibly needed

% use appendices with more than one appendix
% then use \section to start each appendix
% you must declare a \section before using any
% \subsection or using \label (\appendices by itself
% starts a section numbered zero.)
%

\appendices
%\section{Algorithm~\ref{alg0}}
\section*{}
\begin{algorithm}              % enter the algorithm environment
\caption{ROSS phase I: cluster head determination and initial cluster formation for CR node $i$}          % give the algorithm a caption
\label{alg0} 
\DontPrintSemicolon
\SetAlgoLined
\KwIn{$d_j, g_j, j\in \text{Nb}_i\setminus \Lambda$, $\Lambda$ means cluster heads. Empty sets $\tau_1,\tau_2$}
\KwResult{Returning 1 means $i$ is cluster head, then $d_j$ is set to 0, $j\in \text{Nb}_i\setminus \Lambda$. returning 0 means $i$ is not cluster head.}

\If{$\nexists j\in \text{Nb}_i\setminus \Lambda$, such that $d_i \geq d_j$}{
	return 1;
	}
\eIf{$\exists j\in \text{Nb}_i\setminus \Lambda$, such that $d_i > d_j$}{
	return 0;}{
	\If{$\nexists j\in \text{Nb}_i\setminus \Lambda$, such that $d_j == d_i$}{
	$\tau_1 \leftarrow j$
	}
}
\If{$\nexists j\in \tau_1$, such that $g_i \leq g_j$}{
	return 1;
	}
\eIf{$\exists j\in \tau_1$, such that $g_i < g_j$}{
	return 0;
	}
	{\If{$\nexists j\in \tau_1$, such that $g_j == g_i$}{
		$\tau_2\leftarrow j$
		}
	}
\If{$\texttt{ID}_i$ is smaller than any $\texttt{ID}_j$, $j\in \tau_2\setminus i$}{
	return 1;
	}
	{return 0;
	}
\end{algorithm}

% you can choose not to have a title for an appendix
% if you want by leaving the argument blank
%\section*{Algorithm~\ref{alg_size_control_available_CCC}}
\section*{}
\begin{algorithm}               % enter the algorithm environment
\caption{ROSS phase I: cluster head guarantees the availability of CC (start from line 1) / cluster size control (start from line 2)}          % give the algorithm a caption
\label{alg_size_control_available_CCC}
\DontPrintSemicolon
\SetAlgoLined
\KwIn{Cluster C, empty sets $\tau_1, \tau_2$}
\KwOut{Cluster C has at least one CC, or satisfies the requirement on cluster size}
%\tcc*[r]{When to guarantee available CCs, execute from line 1, when to control cluster size, execute from line 2}
\While {$K_C =\emptyset$} {
\While{$|C|> t\cdot \delta$}{
	%calculate $\lambda = \min_{i\in C, i\neq H_C}(|K_{H_C}\cap K_i|)$;\\
	\eIf{$\exists$ only one $i\in C\setminus H_C$, $i = \argmin(|K_{H_C}\cap K_i|)$}{
			$C=C\setminus i$;
		}{
				$\exists$ multiple $i$ which satisfies $i = \argmin(|K_{H_C}\cap K_i|)$;\\ $\tau_1\leftarrow i$;		
		}
		
	\eIf{$\exists$ only one $i\in \tau_1$, $i = \argmax(|\cap_{j\in C\setminus i} K_j|-|\cap_{j\in C} K_j|)$}{
		$C=C\setminus i$;
		}{
			%$\exists$ multiple $i$ which satisfies $i = \argmax(|\cap_{j\in C\setminus i} K_j|-|\cap_{j\in C} K_j|)$;\\
			$C=C\setminus i$, where $i = \argmin_{i\in \tau_1} \texttt{ID}_i $
			%$\texttt{ID}_i$ is smaller than any $\texttt{ID}_j$, $j\in \tau_2\setminus i$;
	}
}
}
\end{algorithm}

\section*{}
\begin{algorithm}               % enter the algorithm environment
\caption{Debatable node $i$ decides its affiliation in phase II of ROSS}
%, chooses one claiming cluster to stay and leaves all the other claiming clusters}          % give the algorithm a caption,  cluster to settle
\label{alg4}
\DontPrintSemicolon
\SetAlgoLined
\KwIn{all claiming clusters $C\in S_i$}
\KwOut{one cluster $C\in S_i$, node $i$ notifies all its claiming clusters in $S_i$ about its affiliation decision.
}
\While{$i$ has not chosen the cluster, or $i$ has joined cluster $\tilde{C}$, but $\exists C'\in S_i, C'\neq \tilde{C}$, which has $|K(C'\setminus i)|-|K(C')|<|K(C\setminus i)|-|K(C)|$}{
	\eIf{$\exists$ only one $C\in S_i$, $C = \argmin(|K(C\setminus i)| - |K(C)|)$}{
			return $C$;
		}{
				$\exists$ multiple $C\in S_i$ which satisfies $C = \argmin(|K(C\setminus i)| - |K(C)|)$;\\ 
				$\tau_1\leftarrow C$;		
		}
	\eIf{$\exists$ only one $C\in \tau_1$, $C = \argmax(K_{h_C}\cap K_i)$}{
			return $C$;
		}{
				$\exists$ multiple $C\in S_i$ which satisfies $C = \argmax(K_{h_C}\cap K_i)$;\\ 
				$\tau_2\leftarrow C$;		
		}
	\eIf{$\exists$ only one $C\in \tau_2$, $C = \argmin|C|)$}{
			return $C$;
		}{
				return $\argmin_{C\in \tau_2}h_C$;\\
		}
		}
\end{algorithm}

% use section* for acknowledgment

\section*{Proof of Theorem~\ref{clustering:theorem}}
\label{proof_clustering:theorem}
\begin{proof}
%Note that the formed cluster can be a singleton cluster, \ie cluster size is 1.
We consider a CRN which can be represented as a connected graph.
% (COMMENT: Why is it necessarily connected? No special case where the graph has two components? ANSWER(Di): It is possible for the graph to have two components. My point it, when theorem works in one component, it works in the complete graph.)
To simplify the discussion, we assume the secondary users have unique individual connectivity degrees. 
Each user has an identical ID and a neighborhood connectivity degree.
This assumption is fair as the neighborhood connectivity degrees and node ID are used to break ties in Algorithm~\ref{alg0}, when the individual connectivity degrees are unique, it is not necessary to use the former two metrics. 
%(COMMENT: What do you mean by breaking ties?? I don't see why your argument holds. ANSWER(DI): according to Algorithm 1, to decide whether node $i$ is a cluster head, the comparision on $i$'s certain degrees with its neighbors needs to be made. It is possible that $i$ has the same degree with some neighbors, then we need ID to decide whether $i$ is cluster head.)

%we assume every node has at least one neighbour.
For the sake of contradiction, let us assume there exist some secondary user $\alpha$ which is not included into any cluster.
Then there is at least one node $\beta\in \text{Nb}_\alpha$ such that $d_{\alpha} > d_{\beta}$. 
According to Algorithm~\ref{alg0}, $\delta$ is not included in any clusters, because otherwise $d_{\beta} = M$, a large positive integer.
Now, we distinguish between two cases: 
%Otherwise, node $\alpha$ is eligible to form a cluster. COMMENT: unnecessary sentence!
If $\beta$ becomes cluster head, node $\alpha$ is included, the assumption is not true.
If $\beta$ is not a cluster head, then $\beta$ is not in any cluster, we can repeat the previous analysis made on node $\alpha$, and deduce that node $\beta$ has at least one neighbouring node $\gamma$ with $d_{\gamma} < d_{\beta}$.
% (COMMENT: $\beta$ not being a cluster head and $\beta$ not belonging to any cluster is not the same statement!!! ANSWER(Di): I add some more words)
Till now, when there is no cluster head identified, the unclustered nodes, \ie $\alpha$, $\beta$ form a linked list, where their connectivity degrees monotonically decrease.
But this list will not continue to grow, because the minimum individual connectivity degree is zero, and the length of this list is upper bounded by the total number of nodes in the CRN.
An example of the formed node series is shown as Figure~\ref{lemma1}.

\begin{figure}[ht!]
  \centering
\includegraphics[width=0.6\linewidth]{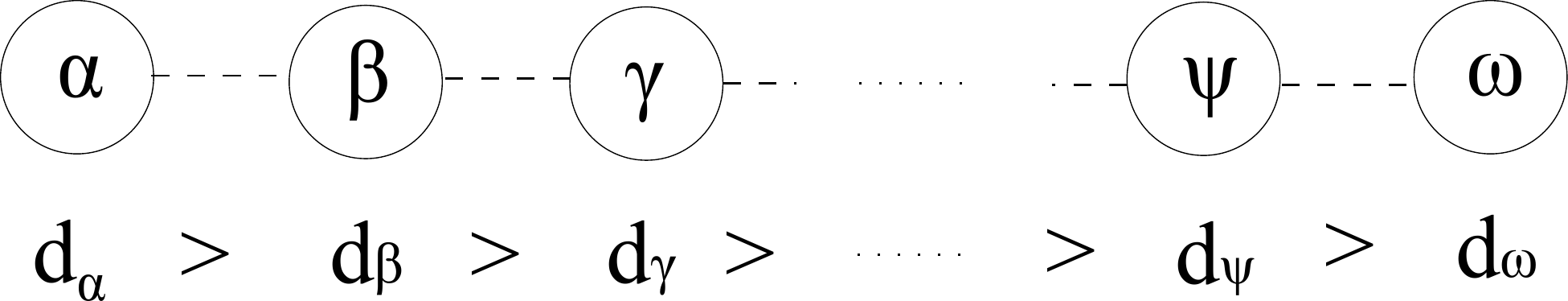}
	\caption{The node series discussed in the proof of Theorem~\ref{clustering:theorem}, the deduction begins from node $\alpha$}
	\label{lemma1}
\end{figure}

In this example, node $\omega$ is at the tail of the list.
As $\omega$ does not have neighboring nodes with lower individual connectivity degree, $\omega$ becomes a cluster head.
Then $\omega$ incorporates all its one-hop neighbours (here we assume that every newly formed cluster has common channels), including the nodes which precede $\omega$ in the list.
The nodes which join a cluster set their individual connection degrees to $M$, which enables the node immediately precede in the list to become a cluster head.
In this way, cluster heads are generated from the tail of list to the head of the list, and all the nodes in the list are in at least one cluster, which contradicts the assumption that $\alpha$ is not included in any cluster.

If we see a secondary user \textit{becoming a cluster head}, or \textit{becoming a cluster member} as one step, as the length of the list of secondary users is not larger than $N$, there are $N$ steps for this scenario to form the initial clusters.

%we know that within at most $N$ steps, all the nodes belong to certain clusters. (COMMENT: No, we don't! Where was that shown?)
\end{proof}

\section*{Proof of Theorem~\ref{theorem1}}
\label{proof_theorem1}
\begin{proof}
To prove the robust clustering problem is NP-hard, we reduce the \textit{maximum weighted k-set packing problem}, which is NP-hard when $k\geqslant 3$~\cite{Computers_a_Intractability}, to the the robust clustering problem to show the latter is at least as hard as the former.
Given a collection of sets of cardinality at most $k$ with weights for each set, the maximum weighted packing problem is that of finding a collection of disjoint sets of maximum total weight.
The decision version of weighted $k$-set packing problem is,
\begin{mydef}
\label{def_kset_packing}
Given a finite set $\mathcal{G}$ of non-negative integers where $\mathcal{G} \subsetneq \mathbb{N}$, and a collection of sets $\mathcal{Q}=\{S_1,S_2,\cdots,S_m\}$ where $S_i \subseteq \mathcal{G}$ and $\max(S_i)\geq 3$ for $1 \leq i \leq m$.
Every set $S$ in $\mathcal{Q}$ has a weight $\omega(S) \in \mathbb{R}$. 
The problem is to find a collection $\mathcal{S} \subseteq \mathcal{Q}$ such that $\mathcal{S}$ contains only pairwise disjoint sets and the total weight of the sets in $\mathcal{S}$ is greater than a given positive number $\lambda$, i.e., $\sum_{S \in \mathcal{S}} \omega(S) > \lambda$.
\end{mydef}

We assume the weights of sets are positive integers.
Then we will show that any instance $\mathcal{I}$ of a weighted $k$-set packing problem, \ie a collection of sets, can be transformed to a clusters formation for a CRN.
W.l.o.g. let set $\mathcal{G} = \{ 1, \ldots , N \}$.
The polynomial algorithm $\sigma$ consists of three steps.

\begin{itemize}

%\item In the first step, we transform the instance $\mathcal{I}$ to $\mathcal{I'}$ by duplicating the elements of the sets in $\mathcal{I}$.
%In particular, as to each element in set $S\in \mathcal{I}$, we duplicate the element and assign a new index which is N bigger than the original index.
%The weight of these sets remain unchanged.
%The purpose of this transformation is to eliminate the set in $\mathcal{I}$, which has only one element.

\item First, the sets in the instance $\mathcal{I}$ are mapped sequentially to the clusters of CR nodes on a two-dimensional Euclidean plane, where the CR user ID is identical with the corresponding element's index.

\item Second, for each mapped cluster $C$, we assign the channels for the nodes in $C$ so that $|K(C)|$ equals to the $\omega(S)$.
We can simply assign the first $|K(C)|$ channels to each CR node in $C$, without considering the possible mismatch when the same CR node appears in different clusters and is assigned with different channels.

\end{itemize}
The number of steps is dependent on $\mathcal{I}$, which is between 1 and $N^2$

%We have found the polynomial algorithm $\sigma$ which transforms an instance of weighted $k$-set packing to an instance of the robust clustering problem in CRN.
Assume we have a robust clustering black box which can check whether the clustering instance meets the requirement, \ie clusters are not overlapping and the total sum of CCs exceed $\lambda$ or not.
If \textit{yes} is said, then the total weight of the corresponding instance of the maximum weighted k-set packing problem is greater than $\lambda$.
If the black box said \textit{no}, either due to overlapping clusters, or the sum of CCs over all clusters is smaller than $\lambda$, the corresponding instance of the packing problem is not an solution.

% The direction is wrong!!
%When $\mathcal{I}$ is not an instance for weighted $k$-set packing problem due to the existence of joint sets, the corresponding clustering instance is not a successful cluster partition for the robust clustering problem, as there are overlapped clusters.
%%
%When the sets in an instance $\mathcal{I}$ for weighted $k$-set packing are disjoint, the sum of weights is identical to the total number of the CCs in the CRN which are mapped from $\mathcal{I'}$.
%Thus, when a instance $\mathcal{I}$ for $k$-set packing problem is true (or false), \ie the sum of weights is greater than $\lambda$, then in the CRN which is mapped from $\mathcal{I'}$, the sum of the numbers of CCs of the clusters is greater (or smaller) than $\lambda$.

Hence, the weighted $k$-set packing can be reduced to the robust clustering problem in CRN, then the latter problem is of NP-hard.
%An example of the reduction is shown in Table~\ref{no_hard_proof_instance}.

%\begin{table}[h!]
%     \begin{center}
%     \begin{tabular}{ C{3cm}  C{4.5cm} }
%     \toprule
%      $\mathcal{N}$  & $\{0, 1,2,3,4,5,6, 7, 8, 9\}$\\ 
%    \cmidrule(r){1-1}\cmidrule(lr){2-2}     
%      $\mathcal{Q}$  & $\{ \{1\}, \{1, 5\}, \{1,2,4\}, \{2,3\}, \{4\} \}$\\ 
%    \cmidrule(r){1-1}\cmidrule(lr){2-2}     
%      Instance for Weighted $k$-set packing  & $\{ \{1\}, \{2,3\}, \{4\} \}$, weights are 2, 3, 4\\ 
%    \cmidrule(r){1-1}\cmidrule(lr){2-2}
%      Instance with dummy elements (step 1)
%      & 
%		$\{ \, \{1,11\},\{2,12, 3, 13\}, \{ 4,14\} \, \}$
%      \\    \cmidrule(r){1-1}\cmidrule(lr){2-2}
%	clusters mapped from $\mathcal{I}$  (step 2)
%	  &
%	  \begin{minipage}{.3\textwidth}	
%      \includegraphics[width=0.6\linewidth]{np_hard_proof_step_2_3.pdf}
%    \end{minipage}
%          \\    \cmidrule(r){1-1}\cmidrule(lr){2-2}
%	formation of clusters (step 3)
%	  &
%	  \begin{minipage}{.3\textwidth}	
%      \includegraphics[width=0.6\linewidth]{np_hard_proof_step4.pdf}
%    \end{minipage}
%	  \\       \bottomrule
%      \end{tabular}
%      \caption{}
%      \label{no_hard_proof_instance}
%      \end{center}
%      \end{table}

\end{proof}

%\section*{Acknowledgment}
%
%The authors would like to thank xxxx

% Can use something like this to put references on a page
% by themselves when using endfloat and the captionsoff option.
\ifCLASSOPTIONcaptionsoff
  \newpage
\fi

% trigger a \newpage just before the given reference
% number - used to balance the columns on the last page
% adjust value as needed - may need to be readjusted if
% the document is modified later
%\IEEEtriggeratref{8}
% The "triggered" command can be changed if desired:
%\IEEEtriggercmd{\enlargethispage{-5in}}

% references section

% can use a bibliography generated by BibTeX as a .bbl file
% BibTeX documentation can be easily obtained at:
% http://mirror.ctan.org/biblio/bibtex/contrib/doc/
% The IEEEtran BibTeX style support page is at:
% http://www.michaelshell.org/tex/ieeetran/bibtex/
%\bibliographystyle{IEEEtran}
% argument is your BibTeX string definitions and bibliography database(s)
%\bibliography{IEEEabrv,../bib/paper}
%
% <OR> manually copy in the resultant .bbl file
% set second argument of \begin to the number of references
% (used to reserve space for the reference number labels box)

% You can push biographies down or up by placing
% a \vfill before or after them. The appropriate
% use of \vfill depends on what kind of text is
% on the last page and whether or not the columns
% are being equalized.

%\vfill

% Can be used to pull up biographies so that the bottom of the last one
% is flush with the other column.
%\enlargethispage{-5in}
\bibliographystyle{IEEEtran}
\bibliography{myrefs}

% biography section
% 
% If you have an EPS/PDF photo (graphicx package needed) extra braces are
% needed around the contents of the optional argument to biography to prevent
% the LaTeX parser from getting confused when it sees the complicated
% \includegraphics command within an optional argument. (You could create
% your own custom macro containing the \includegraphics command to make things
% simpler here.)
%\begin{IEEEbiography}[{\includegraphics[width=1in,height=1.25in,clip,keepaspectratio]{mshell}}]{Michael Shell}
% or if you just want to reserve a space for a photo:
%IEEEbiographynophoto

\end{document}